\documentclass[iop]{emulateapj}
\usepackage{natbib} 
\usepackage{psfig,xspace} 
\shorttitle{X-ray spectrum of Circinus}
\shortauthors{P. Ar\'evalo et al.}

\def\mnras{MNRAS}
\def\apj{ApJ}
\def\aap{A\&A}
\def\apjl{ApJL}
\def\nat{Nature} 
\def\apjs{ApJS}

\def\aj{AJ}

\def\xmm{{\sl XMM-Newton}}
\def\nustar{{\sl NuSTAR}}
\def\chandra{{\sl Chandra}}
\def\swift{{\sl Swift}}
\def\hetg{{HETG}}
\def\heg{{HEG}}
\def\meg{{MEG}}
\def\rosat{{\sl ROSAT}}
\def\sax{{\sl BeppoSAX}\xspace}
\def\suz{{\sl Suzaku}}
\def\asca{{\sl ASCA}}

\def\msun{M$_\odot$}
\def\mekal{{\tt Mekal}}
\def\mytorus{{\tt MYTorus}}
\def\torus{{\tt Torus}}
\def\pexrav{{\tt pexrav}}
\def\pexmon{{\tt pexmon}}
\def\torus{{\tt Torus}}
\def\apec{{\tt APEC}}
\def\redchi2{$\chi^2_{\rm red}$}
\def\flux{erg s$^{-1}$cm$^{-2}$}
\def\norm{photons keV$^{-1}$ s$^{-1}$cm$^{-2}$}
\def\ergs{erg s$^{-1}$}
\def\nh{$N_{\rm H}$}
\def\cm{cm$^{-2}$}
\def\kms{km s$^{-1}$}

\begin{document}

\title{The 2--79~keV X-ray Spectrum of the Circinus Galaxy with NuSTAR, XMM-Newton and Chandra: a Fully Compton-Thick AGN}  
\author{P.\ Ar\'evalo\altaffilmark{1,2}, F.~E.~Bauer\altaffilmark{1,3,4}, S.~Puccetti\altaffilmark{5,6}, D.~J.~Walton\altaffilmark{7}, M.~Koss\altaffilmark{8}, S.~E.~Boggs\altaffilmark{9}, W.~N.~Brandt\altaffilmark{10,11},  M. Brightman\altaffilmark{12}, F.~E.~Christensen\altaffilmark{13}, A. Comastri\altaffilmark{14}, W. W. Craig\altaffilmark{9,15}, F. Fuerst\altaffilmark{7}, P. Gandhi\altaffilmark{16}, B. W. Grefenstette\altaffilmark{7}, C.\  J. Hailey\altaffilmark{17}, F. A. Harrison\altaffilmark{7},  B.~Luo\altaffilmark{10,11}, G.~Madejski\altaffilmark{18}, K.~K.~Madsen\altaffilmark{7}, A.~Marinucci\altaffilmark{19}, G.~Matt\altaffilmark{19}, C.~Saez\altaffilmark{1,20}, D.~Stern\altaffilmark{21}, M.~Stuhlinger\altaffilmark{22}, E.~Treister\altaffilmark{23}, C.~M.~Urry\altaffilmark{24}, W.~W.~Zhang\altaffilmark{25}}

\altaffiltext{1}{Instituto de Astrof\'{\i}sica, Facultad de F\'{i}sica, Pontificia Universidad Cat\'{o}lica de Chile, 306, Santiago 22, Chile}
\altaffiltext{2}{Instituto de F\'isica y Astronom\'ia, Facultad de Ciencias, Universidad de Valpara\'iso, Gran Bretana N¼ 1111, Playa Ancha, Valpara\'iso, Chile.}
\altaffiltext{3}{Millennium Institute of Astrophysics}
\altaffiltext{4}{Space Science Institute, 4750 Walnut Street, Suite 205, Boulder, Colorado 80301}
\altaffiltext{5}{ASDC-ASI, Via del Politecnico, 00133 Roma, ITALY}
\altaffiltext{6}{INAF Osservatorio Astronomico di Roma, via Frascati 33,00040 Monte Porzio Catone (RM), ITALY}
\altaffiltext{7}{Cahill Center for Astronomy and Astrophysics, California Institute of Technology, Pasadena, CA 91125, USA }
\altaffiltext{8}{Institute for Astronomy, Department of Physics, ETH Zurich, Wolfgang-Pauli-Strasse 27, CH-8093 Zurich, Switzerland}
\altaffiltext{9}{Space Sciences Laboratory, University of California, Berkeley, CA 94720, USA}
\altaffiltext{10}{Department of Astronomy and Astrophysics, The Pennsylvania State University, University Park, PA 16802, USA}
\altaffiltext{11}{Institute for Gravitation and the Cosmos, The Pennsylvania State University, University Park, PA 16802, USA}
\altaffiltext{12}{Max-Planck-Institut fŸr extraterrestrische Physik, Giessenbachstrasse 1, D-85748, Garching bei MŸnchen, Germany}
\altaffiltext{13}{Danish Technical University, Lyngby, Denmark}
\altaffiltext{14}{INAF Ð Osservatorio Astronomico di Bologna, Via Ranzani 1, I-40127 Bologna, Italy}
\altaffiltext{15}{Lawrence Livermore National Laboratory, Livermore, CA 94550, USA)}
\altaffiltext{16}{Department of Physics, Durham University, South Road, Durham, DH1 3LE, UK}
\altaffiltext{17}{Columbia Astrophysics Laboratory and Department of Physics, Columbia University, 550 West 120th Street, New York, NY 10027, USA}
\altaffiltext{18}{Kavli Institute for Particle Astrophysics and Cosmology, SLAC National Accelerator Laboratory, Stanford University, 2575 Sand Hill Road M/S 29, Menlo Park, CA 94025, USA}
\altaffiltext{19}{Dipartimento di Matematica e Fisica, Universit\`a degli Studi Roma Tre, via della Vasca Navale 84, 00146 Roma, Italy }
\altaffiltext{20}{Department of Astronomy, University of Maryland, College Park, MD 20742-2421, USA}
\altaffiltext{21}{Jet Propulsion Laboratory, California Institute of Technology, Pasadena, CA 91109, USA}
\altaffiltext{22}{European Space Astronomy Centre (ESAC), ESA, PO Box 78, 28691 Villanueva de la Ca\~nada, Madrid, Spain} 
\altaffiltext{23}{Universidad de Concepci\'on, Departamento de Astronom\'ia, Casilla 160-C, Concepci\'on, Chile}
\altaffiltext{24}{Yale Center for Astronomy and Astrophysics, Physics Department, Yale University, PO Box 208120, New Haven, CT 06520-8120, USA}
\altaffiltext{25}{NASA Goddard Space Flight Center, Greenbelt, MD 20771, USA}

\begin{abstract}

The Circinus galaxy is one of the nearest obscured AGN, making it an ideal target for detailed study.   Combining archival \chandra\ and \xmm\ data with new \nustar\ observations, we model the 2--79~keV spectrum to constrain the primary AGN continuum and to derive physical parameters for the obscuring material.  \chandra 's high angular resolution allows a separation of nuclear and off-nuclear galactic emission.  In the off-nuclear diffuse emission we find signatures of strong cold reflection, including high equivalent-width neutral Fe lines.  This Compton-scattered off-nuclear emission amounts to 18\%\ of the nuclear flux in the Fe line region, but becomes comparable to the nuclear emission above 30 keV.  The new analysis no longer supports a prominent transmitted AGN component in the observed band. We find that the nuclear spectrum is consistent with Compton-scattering by an optically-thick torus, where the intrinsic spectrum is a powerlaw of photon index $\Gamma = $2.2-2.4, the torus has an equatorial column density of $N_{\rm H} = (6-10)\times10^{24}$~\cm\ and the intrinsic AGN 2--10 keV luminosity is $(2.3-5.1)\times 10^{42}$~\ergs. These values place Circinus along the same relations as unobscured AGN in accretion rate-vs-$\Gamma$ and $L_X$-vs-$L_{IR}$ phase space. \nustar 's high sensitivity and low background allow us to study the short time-scale variability of Circinus at X-ray energies above 10~keV for the first time. The lack of detected variability favors a Compton-thick absorber, in line with the the spectral fitting results.  
\end{abstract}


\section{Introduction}   
The Circinus galaxy, hereafter referred to as Circinus, at a distance of $\sim 4.2$ Mpc \citep{freeman77}, contains one of the nearest active galactic nuclei (AGN), and shows signatures of both starburst activity and black hole accretion. The presence of an AGN was initially inferred from high-ionization lines observed in the optical and IR bands. \citet{moorwood84} classified the galaxy as a Seyfert 2 based on the high [NII]/H$\alpha$ ratio and narrow emission lines. This identification was confirmed by the energy input required to produce the observed highly ionized coronal lines \citep{oliva94}. The Seyfert nucleus is obscured from direct view, but the broad line region is visible in reflection, through polarized light \citep{oliva98}.

\citet{greenhill03} detected H$_2$O maser emission consistent with a warped edge-on disc as close in as 0.1 pc from the nucleus. With a line-of-sight velocity of the innermost disc of 260 \kms , this observation implies a black hole mass of $1.5\times 10^{6}$ \msun , consistent with the dynamical upper limit of  $4\times 10^{6}$ \msun\ obtained by \citet{maiolino98}. The bolometric luminosity, $L_{\rm bol}=4\times 10^{43}$ \ergs , has been estimated from the mid-IR nuclear spectrum, which is identified as reprocessed AGN emission \citep{moorwood96}. Combined with the black hole mass estimate, this luminosity equals 20\% of the Eddington luminosity. Therefore, Circinus contains a Seyfert nucleus, { similar in mass and accretion rate to other well known local Seyferts, in particular NGC 4051 \citep{revmap, woo}, but} that is highly obscured from our line of sight. The obscuring material does not cover the entire central engine, as evidenced by the prominent ionization cone observable in [O III] line emission \citep{marconi94}.   

Circinus was identified as an X-ray source for the first time by \citet{brinkmann94} using data from the \rosat\ All Sky Survey. \citet{matt96} discovered strong Fe K$\alpha$ emission with an equivalent width of 2~keV and a hard continuum from a pointed \asca\ observation. This observation revealed strong Compton reflection in this source, produced by an AGN heavily obscured by cold circumnuclear matter, consistent with the Seyfert 2 nature of its optical spectrum.  \citet{matt99} and  \citet{guainazzi99} also found an excess of X-ray emission at higher energies using the \sax\ PDS instrument, interpreted as Compton-scattered and transmitted AGN continuum through clouds of high column density, identified with the molecular torus.  { The Compton shoulder of the Fe lines proves that the cold reflection is produced by Compton thick matter, supporting a scenario where the scatterer and obscurer are the same structure \citep{bianchi02,molendi03}.}
The width of the neutral Fe lines supports the interpretation of the torus, or other structures further out, as the optically thick reflecting material.  \citet{shu11} measured a  FWHM of 1710 \kms\  for the Fe K$\alpha$ line, showing lower velocities than the broad line region, whose H$\alpha$ line observed in polarized light has a FWHM of 3300 \kms\  \citep{oliva98}.

\citet{bianchi01} used \asca\ and \sax\ spectra to determine the origin of the strong emission lines that dominate the X-ray spectrum below 2~keV. Spectral fitting showed that these lines are the result of photoionization rather than collisional ionization, suggesting that they are the result of AGN continuum emission shining on cold material. \citet{sambruna01}  modeled the higher resolution \chandra\ grating X-ray spectra of the nucleus of Circinus  and concluded that the soft X-rays are indeed reprocessed nuclear emission from both photo-ionized and photo-excited plasma. Therefore, at least two distinct reflectors are present in Circinus: one neutral and optically thick responsible for the neutral Fe lines and Compton-scattered continuum, and one photo-ionized, optically-thin plasma responsible for the ionized emission lines of lighter elements which dominate the soft X-ray spectra. The soft X-ray emission results were confirmed by  \citet{guainazzi07} who related the photo-ionized emitting plasma to the narrow line region of this AGN and by \citet{massaro06} who showed that the nuclear spectrum actually requires three reflectors since the ionized, optically thin spectrum must be produced by at least two different regions.

A weak jet and low power nuclear emission have been observed in 6--20 cm wavelengths in the nucleus of Circinus, where the jet collimation suggests that the emission originates in the AGN. The radio power of the nuclear component, however, is very small, about $10^{38}$ \ergs\ \citep{elmouttie98} which corresponds to  $10^{-7} L_{\rm Edd}$. The $B$-band luminosity is not directly observable but the ``blue bump" luminosity estimated from the mid-IR spectral fitting is  $4\times10^{43}$ \ergs, or 20\% of $L_{\rm Edd}$. These values place Circinus comfortably in the range of radio-quiet Seyfert galaxies, as shown in the compilation of \citet{sikora07}. This low radio-power and relatively high accretion rate are similar to the properties of average Seyfert galaxies and consistent with stellar-mass black-hole binaries in the soft state \citep[e.g][]{mcclintock06}.

The proximity of this AGN makes it an ideal target to study in detail both for the properties of the obscurer, as well as for the expected contamination by other sources of emission and scattering outside the nucleus. This contaminating X-ray emission can cause confusion when interpreting the spectra of more distant obscured AGN. 

In this paper we bring together data with high spatial and spectral resolution and wide energy coverage to constrain the physical parameters of the obscuring matter and quantify the non-nuclear contributions. In  Sec. \ref{observations} we describe the main features of the observations performed with each observatory and the reduction procedures. In Sec. \ref{nustar} we compare the new \nustar\ spectrum to previous observations in similar energy ranges and check its consistency to spectral models fit to the earlier data. We then exploit the timing capabilities of \nustar\ to produce the first high-energy, high time-resolution lightcurves of Circinus and use the timing properties to further constrain the spectral decomposition. To make an accurate spectral decomposition, we attempt to model all the non-nuclear X-ray sources within the \nustar\ extraction region using the higher angular resolution data of \xmm\ and \chandra . Toward this end, in  Sec. \ref{low_energy}  we analyze the low-energy spectra using \chandra\ to fit separate models to the nuclear spectrum, non-nuclear point sources and the diffuse emission within Circinus. In Sec. \ref{xmm} we compare the spectra of the nuclear and brightest point sources using the two \xmm\ observations of Circinus, which cover the longest time span. This is used to look for spectral and flux evolution and isolate the behavior of the different regions.   Sec. \ref{fits} presents the joint fits of the galactic and nuclear components of the 2--79~keV spectra using all data sets and in Sec. \ref{conclusion} we summarize our findings.  

\section{Observations}
\label{observations}
We analyze X-ray observations of Circinus obtained by the  \nustar , \chandra , \xmm, \swift , \suz\ and \sax\ observatories between 1998 and 2013. Table \ref{obs_summary} summarizes the primary characteristics of each observation used. 

\begin{table}
\caption{Summary of observations. \label{obs_summary}}
\begin{tabular}{ l l l l }
\hline
Observatory&Date& Exposure [ks] &OBSID\\
\hline
\sax & 1998-03-03&26&5004700200\\
\sax &2001-01-07&16.7&5004700200\\
\xmm &2001-08-06&103&0111240101\\
\chandra\ \hetg &2000-06-15& 68.2 &374+62877\\
\chandra\ \hetg &2004-06-02& 55.0& 4770\\
\chandra\ \hetg &2004-11-28& 59.5& 4771\\
\suz &2006-07-21&140 &701036010\\
\chandra\ \hetg &2008-12-08& 19.7& 10226\\
\chandra\ \hetg &2008-12-15 &103.2& 10223\\
\chandra\ \hetg &2008-12-18& 20.6& 10832\\
\chandra\ \hetg &2008-12-22 &29.0& 10833\\
\chandra\ \hetg &2008-12-23 &77.2 &10224\\
\chandra\ \hetg &2008-12-24 &27.8 &10844\\
\chandra\ \hetg &2008-12-26 &68.0 &10225\\
\chandra\ \hetg &2008-12-27 &37.4 &10842\\
\chandra\ \hetg &2008-12-29 &57.3 &10843\\
\chandra\ \hetg &2009-03-01 &18.1 &10873\\
\chandra\ \hetg &2009-03-04 &16.5 &10872\\
\chandra\ \hetg &2009-03-03 &13.9 &10850\\
\chandra\  ACIS-S &2010-12-17 &152.4 &12823\\
\chandra\  ACIS-S &2010-12-24 &38.9 &12824\\
\nustar&2013-01-25&53& 60002039002\\
\nustar&2013-02-02 & 18& 30002038002\\
\nustar&2013-02-03 &40 &30002038004 \\
\xmm &2013-02-03&57&0701981001\\
\nustar&2013-02-05 & 36&30002038006\\
\hline
\end{tabular}

\end{table}

\subsection{\nustar}
\begin{figure*}
\psfig{file=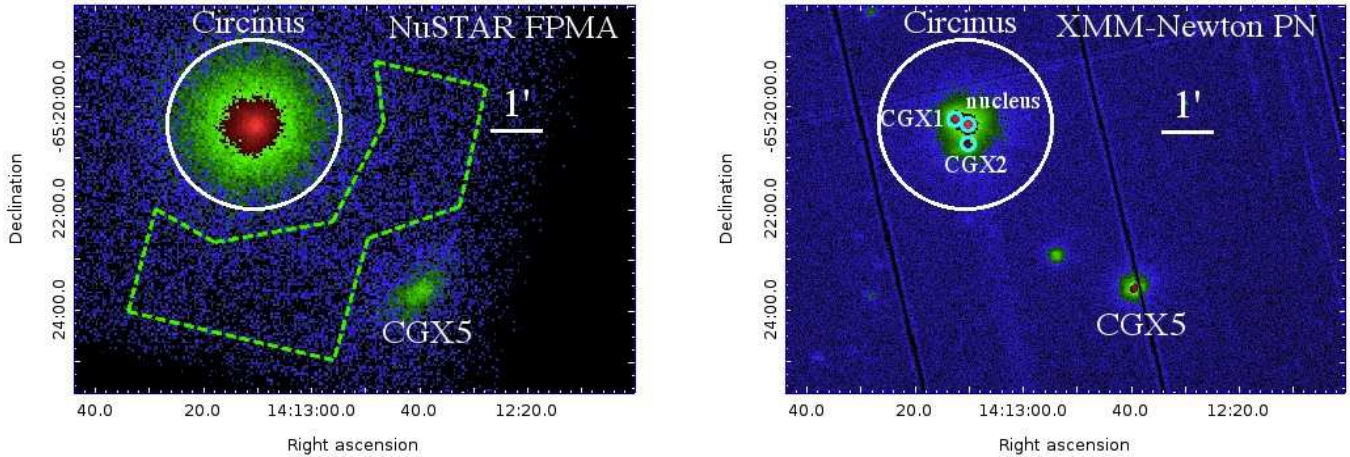,width=18.0cm,angle=0}
\caption{ Left: \nustar\ FPMA image of Circinus in the 3--79~keV range. CGX5 is visible in the lower right corner. The white and green lines mark the source and background extraction. Right: \xmm\ pn image in the 0.2--10~keV range of the same area. The same source source extraction region, marked by the white circle, was used for both observatories. In the pn image the nucleus plus CGX1 and CGX2 are clearly visible within this source region.  The background was extracted from a source free region on the same chip as the source, outside this figure. Both images correspond to the on-axis observations.  }
\label{nustar_xmm_fig}
\end{figure*}

The \nustar\ observatory \citep{harrison13} performed four observations of Circinus between 2013-01-25 and 2013-02-05. The first observation targeted the central AGN on-axis, while the three later observations targeted the ultra-luminous X-ray binary Circinus ULX5 lying 4\arcmin\ to the southwest of the nucleus (\citealt{walton13}, see Fig. \ref{nustar_xmm_fig}) such that the AGN was significantly off-axis. Basic observational details are listed in Table 1. The data were reduced using the standard pipeline ({\sc nupipeline}) from the \nustar\ Data Analysis Software (v1.2.0) within the HEASoft package (v6.14), in combination with CALDB v20130509. Unfiltered event lists were screened to reduce internal background at high energies via standard depth corrections and removal of SAA passages. {\sc xselect} was used to extract data products from the cleaned event lists for both focal plane modules (FPMA and FPMB). Spectra and lightcurves for the AGN were generated using 100\arcsec\ radius apertures, while backgrounds were estimated from blank regions free from contaminating point sources on the same detector (see Fig.  \ref{nustar_xmm_fig}). As a cross-check of our `local' background, we also generated a model of the expected background for each FPM for our adopted aperture with {\sc nuskybgd} \citep{wik14}. {\sc nuskybgd} uses several user-defined background regions spread over all four detectors in each FPM of \nustar\ to fit simultaneously the spectral and spatial dependencies for each background component (e.g., instrumental, focused, and unfocused). {Note the angular resolution of \nustar\ (FWHM=18\arcsec ) is similar to that of \xmm\ and much higher than the other high-energy detectors working in the same energy range as \nustar : \sax\ PDS (12\arcmin ), \swift\ BAT (22\arcmin ) and \suz\ PIN ($4^o.5$).  }

\subsection{\xmm}
Circinus was observed with \xmm\ for two long integrations, a 102.5 ks on-axis observation in 2001 and a 57~ks off-axis observation in 2013, targeting CGX5.  The observations were made in the Full Window mode using the medium filter. We processed both data sets using SAS version 13.0.0. The events files were filtered to exclude background flares selected from time ranges where the 10--12~keV count rates in the pn camera exceeded 0.6 counts per second in 2001 and 0.8 counts per second in 2013. The remaining good exposure time is                                                                                                                                                                                                                                                  78.4 ks (70.5 ks live time) for the 2001 observation and 41.2 ks (36.5 ks live time) for the 2013 observation. 

Three bright central point sources are evident in the pn images, corresponding to the nucleus of the galaxy, the X-ray binary (XRB) CGX1 and the supernova (SN) remnant CGX2, both identified by \citet{bauer01}. We extracted spectra separately for each of these point sources using 11\arcsec\ radius circular apertures and for the whole \nustar\ region using a 100\arcsec\ radius aperture centered on the nucleus.  Background regions were selected on the same chip where the sources were located. 

For the \nustar\ aperture, a rectangular background region was selected away from all visible sources. This procedure is equivalent to the \nustar\ data case, where all nuclear and contaminating sources are extracted in a single spectrum and the background is selected away from the galaxy. For the small point source extractions we selected background regions immediately surrounding the sources, in order to isolate their spectra from contaminating nuclear and diffuse emission within Circinus. Since all these objects contribute to the \nustar\ spectrum, possible variability between epochs is examined in Sec. \ref{xmm} to inform the joint fit of the \xmm\ and \nustar\ spectra.

 We constructed response matrices and ancillary response files using the tasks \emph{rmfgen} and \emph{arfgen} for each epoch and extraction region. The spectra were then binned to contain a minimum of 30 counts per bin. 

\subsection{\chandra}
\label{chandra}
Circinus has been observed on several occasions with \chandra\ over the past $\approx$13 yrs in both its standard ACIS-S and ACIS-S +\hetg\ configurations. We employ the ACIS-S spatially resolved imaging and spectra to model the contribution from contaminating sources around the nucleus of Circinus which fall within the \nustar\ extraction region. Unfortunately the nucleus itself in these data is piled up (60\% pileup within 1-2\arcsec), so the nuclear component must be modeled using the \hetg\ grating spectra. The standard ACIS-S configuration provides a spectral resolution of 100--170~eV between 0.4--8~keV, but can suffer from significant photon pile-up for bright sources due to its high spatial resolution and nominal 3.2s frame time. The addition of the \hetg\ provides an option for higher spectral resolution (60--1000 over the energy range 0.4 --10~keV) at the expense of less effective collecting area. The \hetg\ consists of two different grating assemblies, the High Energy Grating (\heg : 0.8--10~keV) and the Medium Energy Grating (\meg : 0.4--8~keV), which simultaneously disperse a fraction of the incident photons from High Resolution Mirror Assembly (HRMA) shells along two dispersion axes offset by 10 degrees to form a narrow X-shaped pattern along the length of the ACIS-S detector. The gratings preferentially absorb soft undispersed photons but let a fraction of higher-energy photons pass through to comprise the \hetg\ 0$^{th}$ order image on ACIS-S.

We retrieved all of the \chandra\ data shown in Table~1 from the \chandra\ Data Archive, and reduced them following standard procedures with the CIAO software (v4.4) and calibration files (CALDB v4.4.6). We reprocessed the data to include the latest calibrations, remove the 0\farcs5 pixel randomization, and correct for charge transfer inefficiency (CTI). We screened the data with the standard ASCA grade selection, exclusion of bad pixels and columns, and intervals of excessively high background (none was found). Analyses were performed on reprocessed \chandra\ data, primarily using CIAO and custom software including ACIS EXTRACT (v2013-04-29; \citealt{broos10}) for ACIS CCD spectra.

\begin{figure*}
\psfig{file=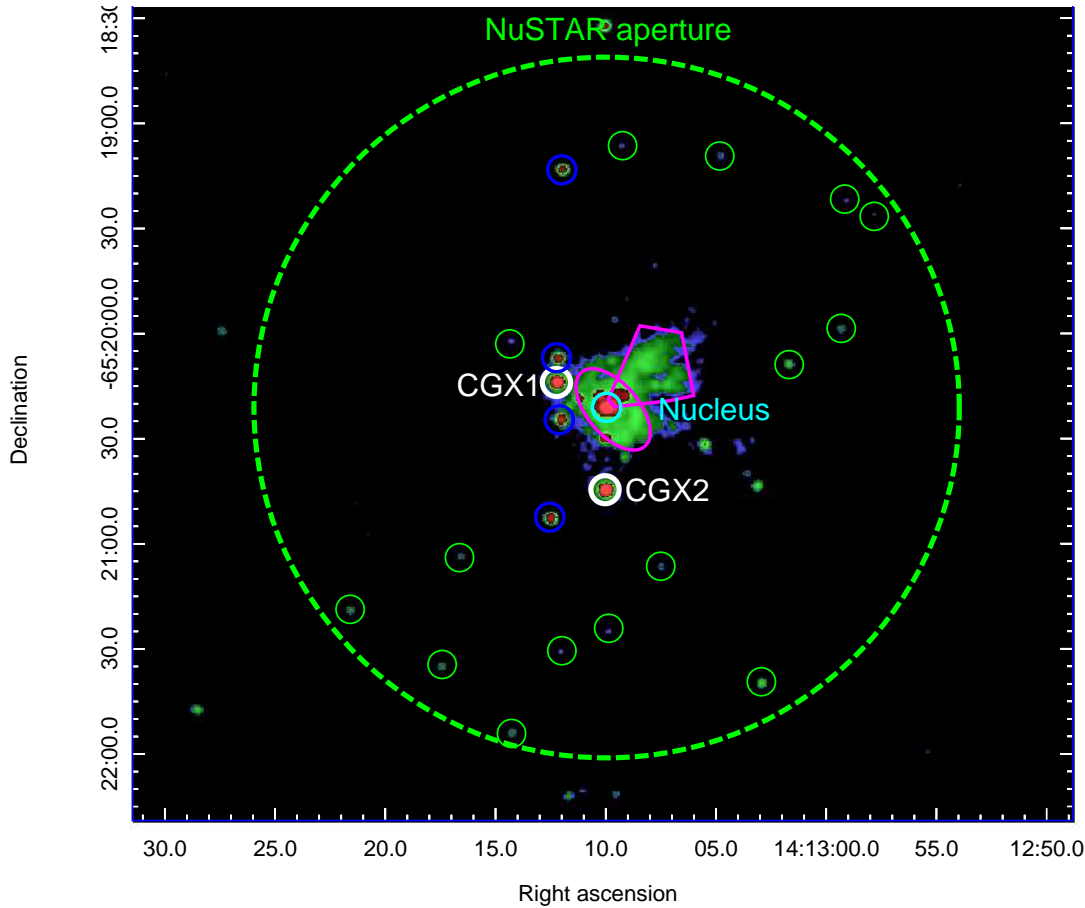,width=16.0cm}
\caption{\chandra\ ACIS image of Circinus.   The green dashed circle shows the extraction region used for the \nustar\ spectra. The point sources modeled individually are marked in white solid lines. The next four brightest point sources are marked in blue and a sample of dim point sources are marked in green. The circles around the point sources represent the  $4\arcsec $ extraction regions used for the \chandra\ spectral modeling. The magenta lines mark the extraction regions for extended emission: an off-center circumnuclear ellipse and the ionization cone. The central region, marked in cyan, was extracted from an annulus that  excludes the piled-up central $2\arcsec$ radius region. \label{chandra_fig}}
\end{figure*}

A \chandra\ 0.4--8~keV image of the region around the Circinus AGN is shown in Fig. \ref{chandra_fig}. Many different sources appear in the low-energy image within the \nustar\ source extraction region, which could potentially contaminate the nuclear emission in the \nustar\ band. Thus to model the nuclear spectrum, we must model all the extra-nuclear sources of emission below 8~keV. To this end, we extracted ACIS-S CCD spectra using standard CIAO tools for several extraction regions as shown in Fig. \ref{chandra_fig} and detailed in Secs. \ref{pointsources} and \ref{diffuse}. For simplicity, we only used OBSID	s 12823 and 12824 to generate and fit the contamination spectra, as these observations comprise 191.2~ks out of 299.7~ks total normal ACIS-S data (64\%), are positioned on-axis, use the entire ACIS-S FOV, and are well-calibrated. The 671.4~ks of \hetg\ 0$^{th}$ order data was not considered due to its significantly larger calibration uncertainties (M. Nowak, private communication).

Since the nucleus is piled up, we modeled it based solely on the higher-order \heg\ and \meg\ spectra. \heg /\meg\ spectral products were extracted from each OBSID using standard CIAO tools using a 6\arcsec\ full-width \heg /\meg\ mask in the cross-dispersion direction centered on the AGN. The moderate energy resolution of the CCD detector ACIS-S was used to separate the overlapping orders of the dispersed spectrum. We combined the plus and minus arms to obtain a composite 1st order spectrum for each grating assembly. We found no obvious variability amongst the multiple \hetg\ observations, and thus combined all of the spectra and averaged the auxiliary response files (ARFs) into a composite spectrum with 671.4~ks equivalent exposure.

\subsection{\suz}
Circinus  was also observed by \textit{Suzaku} \citep{mitsuda07} in 2006 for $\sim$140\,ks. We only consider data from the HXD PIN detector, which covers a similar energy range to \nustar . We reprocessed the unfiltered event files using the {\small HEASOFT} software package (version 6.13) and following the  \textit{Suzaku} Data Reduction Guide.\footnote{http://heasarc.gsfc.nasa.gov/docs/suzaku/analysis/} Since the HXD is a collimating instrument, estimating the background requires individual consideration of the non X-ray instrumental background (NXB) and cosmic X-ray background (CXB). The response and NXB files provided by the \textit{Suzaku} team were downloaded\footnote{http://www.astro.isas.ac.jp/suzaku/analysis/hxd/}; here we use the higher quality `tuned' (Model D) background. Spectral products were generated using the {\small HXDPINXBPI} script, which also simulates the expected contribution from the CXB using the form of \cite{boldt87}. Finally, we rebinned the PIN data so that each energy bin had a minimum S/N of 3, 
sufficient to allow the use of $\chi^{2}$ minimization. The binned spectra cover the range 15--96~keV. Although ULX5 discussed previously is contained in the HXD field-of-view, the nuclear emission strongly dominates at these energies \citep{walton13}.

\subsection{\sax}
Circinus was observed by \sax\ on 1998-03-13 (OBSID=5004700200, and on time$\sim$26 ksec), and on 2001-01-07 (OBSID=5004700200 and on time$\sim$16.7 ksec). The Phoswitch Detector System  (PDS, \citealt{frontera97}) data were calibrated and cleaned using SAXDAS software with the standard method `fixed Rise Time threshold' method for background rejection.  The PDS lightcurves are well known to exhibit spikes on timescales between a fraction of second to a few seconds and usually most counts from spikes are recorded below 30~keV. To screen the PDS data for these spikes we followed the method suggested in the NFI user guide\footnote{http://heasarc.nasa.gov/docs/sax/abc/saxabc/saxabc.html}. The PDS spectra were logarithmically binned in the 15--220~keV energy range in 18 channels.

\subsection{\swift}
Since November 2004, the Burst Alert Telescope (BAT) on board Swift \citep{gehrels04} has been monitoring the hard X-ray sky (14--195~keV). \swift /BAT uses a 5200 cm$^2$ coded-aperture mask above an array of 32768 CdZnTe detectors to produce a wide field of view of 1.4 steradian of the sky.  Since BAT is continuously observing the sky, a new snapshot image is produced every five minutes for a large number of hard X-ray sources due to its wide field of view and large sky coverage.  To study long term source variability we use the publicly available 70 month lightcurves from \swift\ BAT \citep{baumgartner13}.  Because of the large energy range of BAT, we are able to test any underlying energy dependence of the light curve.  We extracted lightcurves containing count rates in eight different energy bands: 14--20, 20--24, 24--35, 35--50, 50--75, 75--100, 100--150, 150--195~keV.  We also use the stacked 70 month spectra of Circinus.

\section{\nustar\ 3--79~keV spectrum}

\label{nustar}
\begin{figure}
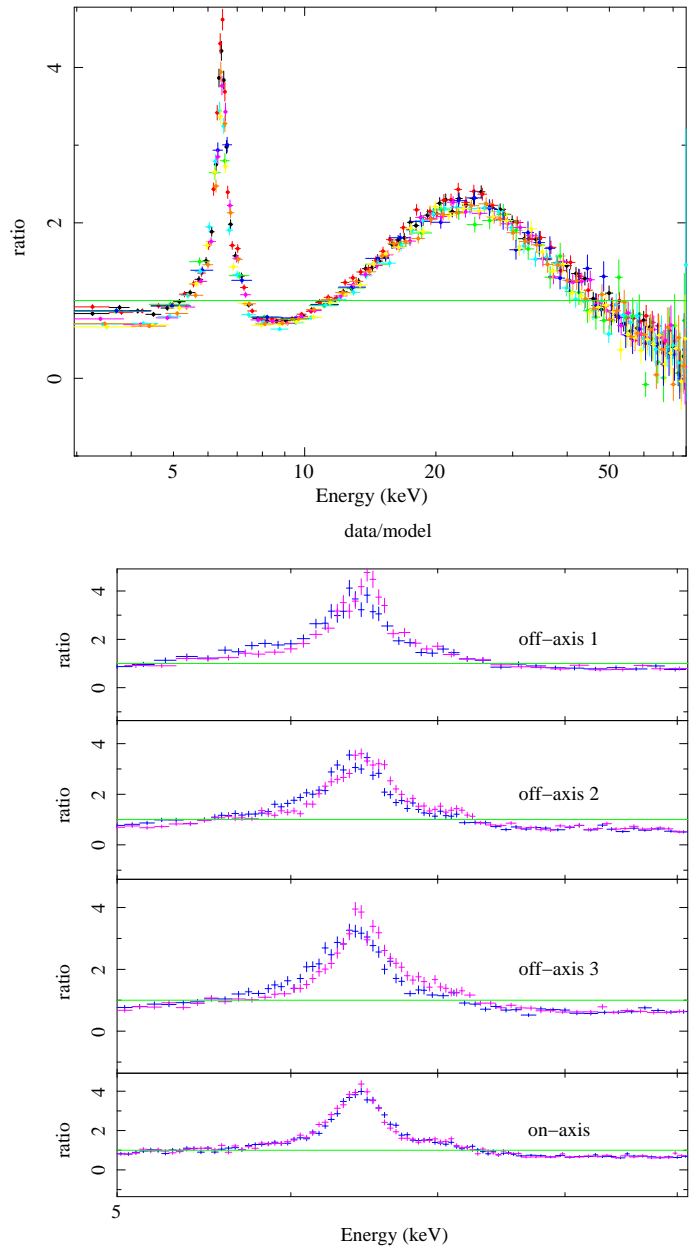

\psfig{file=nustar_all_binned.ps,width=9.0cm,angle=270}
\psfig{file=nustar_Fe.ps,width=9.0cm,angle=270}
\caption{{\emph Top}: ratio of all \nustar\ spectra to a powerlaw of slope 0.89 and normalization of $1.3\times 10^{-3}$\norm\ at 1 keV; both modules and the four exposures produce consistent spectral shapes and normalizations, except for details visible at low energies, where the spectra of FPMA and FPMB of the first two exposures (black, red, green, blue) are slightly higher than the spectra of the last two exposures (cyan, magenta, yellow and orange). {\emph Bottom}: Same spectra as above focusing in on the Fe K line region. Blue points correspond to FPMA and magenta to FPMB, with each panel showing a different OBSID. The on-axis observations produce the most consistent spectra between the modules and have sufficient counts for accurate spectral fitting in this energy range. 
\label{nustar_spectra}}
\end{figure}

We extracted spectra for \nustar\ modules A and B for the on-axis and off-axis observations. The second observation consists of three segments and we initially made a separate spectrum for each. For each epoch, the A and B modules produced largely consistent spectra. The different epochs, however, show differences in the flux at the lowest energies, as shown in Fig. \ref{nustar_spectra}. Below 5~keV, the normalization of the on-axis and first segment of the off-axis observations are significantly higher than the other two segments of the off-axis observation. { The flux of CGX1 is highly variable, as has been shown using spatially resolved \chandra\ data \citep{bauer01}. Since this XRB falls inside the \nustar\ source extraction region, it is possible that a varying XRB flux is producing the observed differences in the \nustar\ spectra. Based on the spectral extraction of CGX1 in the \xmm\ data presented in Sec. \ref{xmm}, CGX1 does appear to be solely responsible for the observed variability.   The differences between the different \nustar\ spectra are consistent with a changing normalization of the absorbed powerlaw component that will be used to model CGX1 below in Sec. \ref{pointsources}. }

 The 3--5~keV flux measured for the four segments is $4.1\times 10^{-12}$\flux\ for the first two and $3.4-3.2\times 10^{-12}$\flux\ for the second and third segments, respectively. The CGX1 flux would have to change by $6.9\times 10^{-13}$ \flux\ to make up for the differences in the spectrum. To account for this difference, in the on-axis \nustar\ observation and the first segment of the off-axis observation the powerlaw normalization of CGX1 should be $6.87\times 10^{-4}$\norm\ at 1 keV higher than in the second and third off-axis observations. This difference is incorporated in the joint fits of Sec. \ref{fits} when combining the spectra of the on-axis \nustar\ observation with the 2013 pn spectrum, which coincides in time with the second segment of the off-axis \nustar\ observation.

Once this powerlaw normalization difference is accounted for, small differences remain around the 6.4~keV emission line, which shows a stronger red wing in the off-axis spectra. Since the energy calibration is less certain at off-axis positions we will use only the on-axis spectra. 
Above 8 keV there are no significant differences between the four different epochs. Fig. \ref{nustar_spectra} shows all the \nustar\ spectra divided by the same powerlaw with $\Gamma =          0.889\pm  0.003$ and norm $=  (1.30  \pm  0.01\times 10^{-3})$ \norm\ at 1~keV, showing that the spectra are highly consistent at most energies and there are no other measurable fluctuations in the spectral shape or flux of the different epochs.  

We compared simultaneous observations of Circinus with \nustar\ and \xmm\ to measure any flux calibration differences. We selected \nustar\ events over the time range where the off-axis observation overlapped with the 2013 \xmm\ observation. The resulting \nustar\ spectra were fitted together with the 2013 pn spectrum using the same spatial extraction region, over the 3--10~keV energy range where the observatories overlap. Fitting both spectra with a powerlaw of equal slope and free normalization gives normalizations of $(5.05  \pm  0.13)\times 10^{-4}$  for pn, $(6.08 \pm 0.20)\times 10^{-4}$ for \nustar\ FPMA and $(6.23\pm  0.20) \times 10^{-4}$ for FPMB. FPM A and B are consistent within the statistical error, while they are both about 20\% above the pn value. For this reason we will multiply the models by a constant factor of 1.2 when fitting them to \nustar\ spectra in the joint fitting described in Sec. \ref{fits}.  We note that this constant cross-calibration factor has changed in the lastest version of the { \nustar} extraction software to a value of $\sim 1$. 

\subsection{Comparison to previous hard X-ray observations}
\label{he_comparison}
\begin{figure}
\psfig{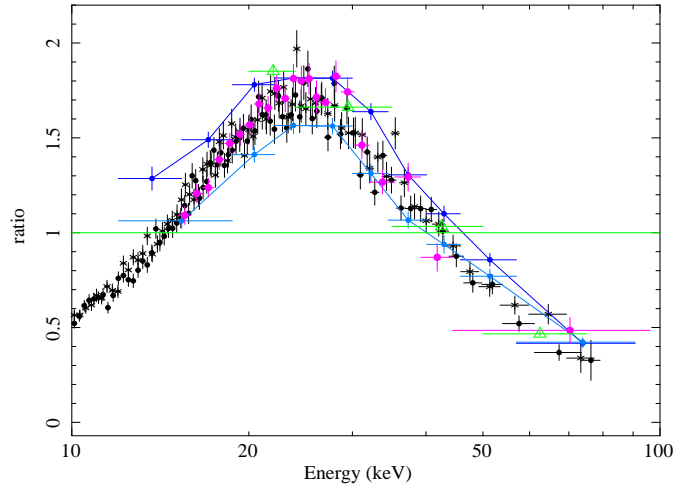}
\caption{Residuals of Circinus spectra taken with different observatories to the same powerlaw model, where only the normalizations were corrected for known mismatches in the response of the instruments.  The on-axis \nustar\ spectra are plotted in black (FPMA in circles, B in stars), two different \sax\ PDS observations are plotted in blue and light blue, \suz\ PIN data are plotted in magenta { and \swift\ BAT in green triangles}. The spectra were binned to similar significance levels. The data are broadly consistent in flux and shape. The difference between the \sax\ spectra is discussed in the text.
\label{he_comp}}
\end{figure}

Circinus had been observed previously above 10~keV by \sax\ in 1998 and 2001 \citep{matt99, bianchi01,bianchi02} and in 2006 by \suz\  \citep{yang09}.  Figure \ref{he_comp} compares the observations from the three satellites, showing the ratio of their spectra to the same powerlaw model,  where only the normalizations were shifted for known mismatches in the response of the instruments. The model normalization has been shifted upward by 20\% for \nustar\ spectra, as discussed above. The \suz\ PIN normalization has been shown to be 16\% too high when compared to the XIS;\footnote{ftp://legacy.gsfc.nasa.gov/suzaku/doc/xrt/suzakumemo-2008-06.pdf} we therefore also corrected its model normalization up by this amount. The PDS normalization has been shown to be too low by 20\%  \citep{grandi97} so we multiplied the model normalization by 0.8 when applied to these data. {  The 70-month average BAT spectrum model was rescaled by a cross-calibration coefficient of 0.95, to match the \nustar /\suz\ spectra.   }
 
 With these corrections to the normalizations, the spectra are very similar, with { \suz\ PIN and \swift\ BAT spectra being fully consistent with  \nustar }. The two \sax\ observations, one lower and one higher than the other spectra, were compared by \citet{bianchi02}. These authors ascribe the large change in flux between observations to variations of CGX1, although the brighter ULX5, 3\arcmin\ away from the nucleus, might also contribute to this spectrum. Even if a 2\arcmin\ extraction region was used, ULX5 is bright enough and variable, such that it could have contaminated the nuclear spectrum given the broad PSF of the \sax\ PDS. The alternative is that the \sax\ PDS observation indeed caught a rare, less obscured, glimpse into the central engine that has not been reproduced since (see Sect. \ref{variability}). Repeating the extraction with a 1\arcmin\ radius reduced the nuclear variation to a level where it was no longer significant  \citep{bianchi02}. ULX5 has a softer spectrum than the nucleus \citep{walton13}, so the ``soft excess" visible in the \sax\ spectra is consistent with this potential contamination.  A further probable cause for this difference in flux between the \sax\ observations is the background subtraction method required by the low-spatial resolution of PDS, whereby a blank field is used to estimate the background contamination. Low-flux sources in this field can cause the background to be over estimated and therefore the source flux to be underestimated. Considering these limitations we do not raise further concerns about the discrepancy between both \sax\ spectra and the \nustar\ spectrum. 
 
 We note that more confidence should be placed in the \nustar\ data, which offer higher quality high-energy spectra than \suz\ PIN and \sax\ PDS due to the focusing nature of the telescope. The large concentration factor leads to greatly reduced internal background levels, higher S/N spectra, and less uncertainty from nearby source contamination. Once convinced that the \nustar\ spectra are compatible with previous observations, we will base our final spectral modeling above 10~keV in Sec. \ref{fits} largely on the \nustar\ data although we also incorporate the \swift\ BAT data to better constrain the highest energies.

\section{Detailed broadband fitting: Spectral components in the 0.8--8~keV range}
\label{low_energy}
The high spatial resolution images of \chandra\ ACIS show a number of point sources as well as extended X-ray emission within the $100\arcsec$ region used to extract the \nustar\ spectrum. In this section we model the nuclear emission and assess the contribution of these contaminants to the nuclear flux. 

\subsection{Nuclear spectrum}
\label{hetg}

\begin{figure}
\psfig{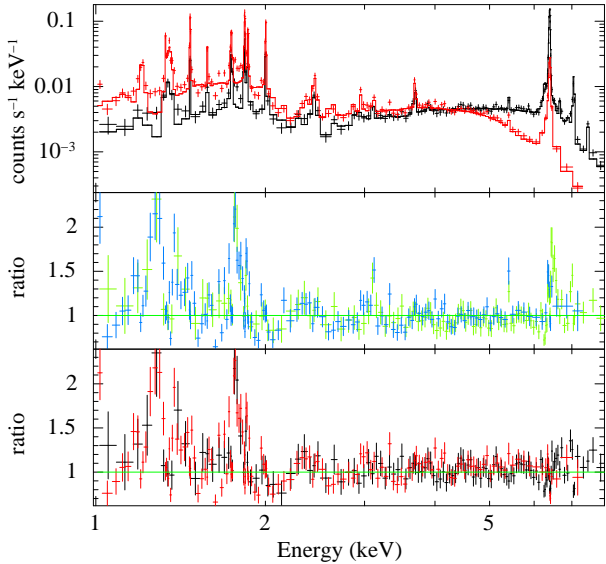}
\caption{The nuclear spectrum: \heg\ (black) and \meg\ (red) spectra extracted from the central 3\arcsec\  radius region, fitted with \mytorus\ scattered+lines plus the Gaussian lines identified in \cite{sambruna01}. The ratio to this model is shown in the middle panel. The spectrum above 2~keV was fit with additional Gaussian lines as described in the text. This final model is fit to the spectra in the top panel and its data/model ratios are shown in the bottom panel.  
\label{grating_spectra}}
\end{figure}

The \heg\ and \meg\ are the only spectra with sufficient spatial resolution to model the central 3\arcsec\ alone, since the ACIS data are strongly piled-up in this region. These spectra, shown in the top panel in Fig. \ref{grating_spectra}, have two distinct features: a forest of narrow emission lines at low energies and, above 2~keV, a hard powerlaw with prominent emission lines at 6.4 and 7.1~keV. The spectrum above 2~keV is characteristic of Compton-scattered and fluorescent lines produced by reflection of the AGN continuum off neutral matter. 

\citet{sambruna01} fitted 60 ks of this \chandra\ \hetg\ observation and interpreted the 1--8~keV spectrum as two distinct reflection components, one from ionized material and one from nearly neutral material.  The soft X-ray emission lines are largely produced by the ionized medium. Line intensities correspond to a photo-ionized and photo-excited medium rather than collisionally ionized gas. { \citet{guainazzi07} analyzed RGS spectra of Circinus below 1~keV finding narrow radiative recombination continua (RRC) which require a low gas temperature of a few eV and indicate that the gas is in photoionization equilibrium.} The neutral reflection found by \citet{sambruna01} is highly concentrated toward the center of the galaxy and is responsible for the strong Fe K$\alpha$ and K$\beta$ lines as well as a hard reflection continuum. 

We initially model the Compton reflection and Fe fluorescent lines with  \mytorus\ scattered and line components. The most prominent emission lines correspond to the K$\alpha$ and K$\beta$ transitions of neutral Fe and are calculated consistently with the Compton-scattered component. These lines are the only ones modeled by  \mytorus , although neutral lines from other elements should also be present at a considerably lower flux level. The cold, photo-ionized medium is simply modeled as the Gaussian lines reported in Table 1 of  \citet{sambruna01}, over a powerlaw representing reflection of the nuclear continuum. The RGS spectrum of Circinus presented by \citet{guainazzi07} shows that lower energy emission is highly dominated by narrow emission lines and therefore is not helpful in constraining the behavior of an underlying scattered or Compton-scattered continuum component. { Given the low count rate of our spectra at the lowest energies, the resolution of our binned-\hetg\ spectra cannot resolve these low-energy lines, so we will restrict our analysis to energies above 1~keV.} The lowest energy line reported by  \citet{sambruna01} is at 1.2~keV. Below this energy many closely separated transitions of Fe XIII through Fe XXIV, plus Na lines are unresolved by our data and resemble continuum emission. We model these collectively as two broad (FWHM=0.01~keV) Gaussian lines at 1 and 1.1~keV. Allowing the rest of the line widths and all normalizations to vary, we obtain a \redchi2 =1.48 for 1287 dof. The overall shape of the spectrum is well reproduced but residuals remain around most of the emission lines. Many of the emission lines reported by \citet{sambruna01} originally modeled several transitions together. Our higher S/N spectrum shows many narrow cores over these moderately broad lines ($\sigma=0.01$~keV).  Our aim is to constrain the continuum components that contribute to higher energies and not to derive the gas properties from line diagnostics. Therefore we fixed the powerlaw slope and normalization from this fit and restrict further analyses to energies above 2~keV where the lines are only a minor contribution to the flux. In the 2--8~keV range this model produces \redchi2 =1.25 for 857 dof.  { We also included lines not reported in  \citet{sambruna01}, including  a 7.4~keV (Ni K$\alpha$, \citealt{molendi03}), $6.55  \pm 0.21$ ~keV (blend of the FeXXV triplet, \citealt{bianchi02}) and  $3.108\pm 0.002$~keV (S XVI K$\beta$), $5.41   \pm  0.01$~keV (Cr K$\alpha$) and $2.18 \pm 0.02$~keV (Si XIII K$\beta$)  $5.885\pm 0.005$ (Cr XXIV K$\alpha$),  modeled as narrow Gaussians with their widths tied}. Repeating the fit with free normalization of these lines and of the \mytorus\ component improves the fit significantly to \redchi2 $=$ 1.04 for 852 dof.

Although this fit is already good, the residuals show striking structure around the Fe K$\alpha$ line, where the line core is under-predicted and the Compton shoulder is over-predicted. { However, as the Fe line emission is spatially extended, the nuclear spectrum has contamination from neighboring energy bins due to incorrect spatial vs. energy assignments along the dispersion direction. We estimate that this spatial/spectral energy separation degeneracy leads to an overall broadening of emission lines by 0.1-0.2 keV (effectively the energy resolution of ACIS-S) and an overestimate of $\sim$20\% in the line flux around 6.4 keV. We included narrow Gaussian lines to simulate this contamination of the strongest lines at 6.4 and 7~keV in the nuclear spectrum}.  Finally fitting the Compton-scattered component together with these lines produces a \redchi2 = 0.97 for 851 dof. 

In addition to the \mytorus\ Compton scattering component, we have included the transmitted AGN continuum using the \mytorus\ zero-order model, since we are looking directly in the direction of the AGN. Given the large optical depth towards the nucleus, this transmitted component does not make a noticeable contribution in the \chandra\ band but might contribute significantly to the spectrum at higher energies. As a first approach we tied all the \mytorus\ components --- Compton-scattered, transmitted spectrum and Fe lines --- to have the same \nh , inclination angle, intrinsic powerlaw slope and normalization, which corresponds to a uniform torus geometry. Since the model already produces a very good fit at energies where the Compton-scattered component dominates, the fit does not improve significantly by freeing these constraints. 

Previously, \citet{molendi03} modeled the 4--12~keV nuclear spectrum using the pn data from the first \xmm\ observation. Their analysis of the Fe line, its Compton shoulder and the depth of the Fe edge indicated that the same Compton thick matter was responsible for the line-of-sight absorption and for the volume-averaged Compton scattering. They therefore identified the reflector with the absorber, possibly corresponding to the molecular torus. The consistency between their parameters pointed to a fairly uniform torus. Their best-fitting model produced an intrinsic powerlaw slope $\Gamma=1.9$ and Fe abundance of 1.2 with respect to solar. In our approach, the Compton-scattered component is modeled with \mytorus\ instead of \pexrav , which does not produce large differences below 10~keV and is not very sensitive to the incident powerlaw slope in this energy range, making this value of $\Gamma$ acceptable in our fits as well. The Fe abundance is also consistent with our findings, since the \mytorus\ Fe lines are calculated using solar abundances and we fix the scaling between Compton-scattered and Fe lines to 1. 

The parameters of the Compton scatterer are not well constrained in this energy range. Allowing only the \nh\ and normalization of \mytorus\ to vary, together with the normalization of the scattered powerlaw, the best-fitting values for \mytorus\ components are \nh =$(9.4\pm7.7)\times 10^ {24}$ \cm , normalization of $0.21\pm0.06$ \norm\ at 1~keV for the Compton-scattered powerlaw, and $(1.5 \pm 0.4)\times 10^{-4}$ \norm\ at 1~keV  for the scattered powerlaw normalization. Higher energy data are necessary to constrain further the Compton scattering component.   

\subsection{Point sources}
\label{pointsources}
After the nucleus, the brightest point sources within  $100\arcsec$ of the nucleus of Circinus are CGX1, a bright X-ray binary, and CGX2, a young supernova remnant. Both were  previously identified and characterized by \citet{bauer01}. We extracted their ACIS 0.5--8~keV spectra using circular regions  4$\arcsec$ in radius.  The next four sources ranked in flux were extracted with the same criteria and combined to produce a spectrum representative of lower luminosity point sources.  A further sample of 15 dimmer point sources were extracted and modeled together. These regions are marked in Fig. \ref{chandra_fig}.

The best known nuclear contaminants are CGX1 and CGX2. We modeled the first one, an XRB, with a simple absorbed powerlaw ({\tt phabs*powerlaw}) which produced a reasonably good fit with \redchi2 = 1.1 for 394 dof. The {\tt phabs} \nh$=(1.03 \pm 0.03)\times10^{22}$~\cm\ is slightly higher than the Galactic value of $3\times 10^{21}$~\cm\ \citep{freeman77},  indicating absorption in the host galaxy. The powerlaw slope is $\Gamma=1.80\pm0.04$ (norm$=2.77\times 10^{-4}$ \norm ), typical of XRB slopes. No strong emission lines are evident. 

The spectrum of CGX2 shows strong emission lines, most clearly He- and H-like Fe at 6.7 and 6.95~keV plus several emission lines in the soft X-rays. This spectrum is produced by the supernova ejecta, composed of hot, optically thin gas of non-solar abundances being ejected at velocities of up to 10,000 \kms . The gas is not in ionization equilibrium and covers a wide range in temperatures which we crudely model using several thermal plasma components. The simplest model that produces a reasonable fit (\redchi2 =1.19 for 415 dof) was three \mekal\ components with temperatures of $0.09\pm 0.02, 1.09\pm0.03$ and $9.15\pm0.38$~keV, abundances in units of Solar abundance of 0.5 (frozen) for the lowest temperature phase and $7.5\pm0.8$ for the other two and redshifts of $0.08\pm0.04$, $-0.0066\pm0.0014$ and $-0.0028\pm0.0018$. All these components are under a layer of cold absorption modeled by {\tt phabs} with \nh $=2.2\times10^{22}$~\cm . 

The remaining point sources were fitted together in groups to improve their signal-to-noise ratio. First the 3rd to 6th brightest were modeled in the same way as CGX1, producing similar fit values. Their combined spectrum are well fitted by an absorbed  powerlaw of slope $\Gamma=1.8$ and column density \nh =$5\times 10^{21}$~\cm . The joint normalization is $1.5\times 10^{-4}$ \norm , a factor of 2 lower than that of CGX1. The next 15 fainter point sources together produced a steeper slope and lower normalization so their contribution is unlikely to be an important source of contamination in the \nustar\ band. 

\begin{figure}
\psfig{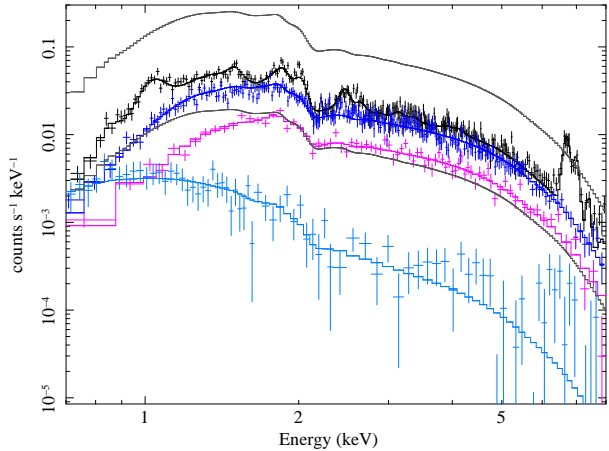}
\caption{\label{ps_spectra} \chandra\ ACIS spectra and models of point sources within the \nustar\ extraction region:  CGX1 (dark blue), CGX2 (black), the next four brightest point sources (magenta) and a sample of 15 dimmer point sources (light blue). All these sources have steep spectra ($\Gamma \sim 1.8$) so they will not contaminate the $>$10keV spectra significantly. The supernova remnant CGX2 contributes the majority of the ionized Fe emission (6.7 and 6.9~keV) in the \nustar\ aperture. The gray lines mark the lower and upper limits of the expected contribution from HMXBs in the entire galaxy, modeled as powerlaws of $\Gamma=1.8$ under a cold absorber of \nh $= 1\times 10^{22}$~\cm .}
\end{figure}

The predicted contribution of high-mass X-ray binaries (HMXBs) was also calculated from the measured star formation rate for the entire Circinus galaxy, which has been estimated to be between  0.2 and 6 M$_\odot$yr$^{-1}$ \citep[e.g.][]{grimm03,roy08}. Using the scaling relation of \citet{lehmer10}  and a distance of 4.2 Mpc,  the predicted 2--10~keV flux produced by HMXBs is $(0.4-5.3)\times 10^{-12}$ \flux .  This level of HMXB contribution is plotted in gray solid lines in Fig. \ref{ps_spectra} along with the various spectra of the resolved point sources highlighted above. Both CGX1 and CGX2 alone fall within the expected range, so together they can represent the majority of the point source contribution to the spectrum below 8.0~keV. We are therefore confident that unresolved point sources will not contaminate significantly more than what is already accounted for in our composite model.

\subsection{Diffuse emission}
\label{diffuse}

Diffuse emission can be seen mainly around the nucleus of Circinus and at the location of the ionization cone, though dimmer emitting regions exist throughout the galaxy. This emission is produced by a combination of hot gas created from the starburst and AGN jets  \citep{mingo12} as well as AGN emission reflected off neutral  \citep[e.g.][]{smithwilson01} and photo-ionized material \citep{sambruna01}.  

We extracted spectra for four distinct regions: the ionization cone region marked  in Fig. \ref{chandra_fig}; a circumnuclear ellipse excluding the central 3$\arcsec$ which corresponds to the nucleus; nuclear emission within 3$\arcsec$ but excluding the innermost 2$\arcsec$ radius, which is significantly piled-up, and all the remaining area enclosed in  the \nustar\ aperture (100$\arcsec$ radius) excluding the diffuse and point source regions discussed above. 

\begin{figure}
\psfig{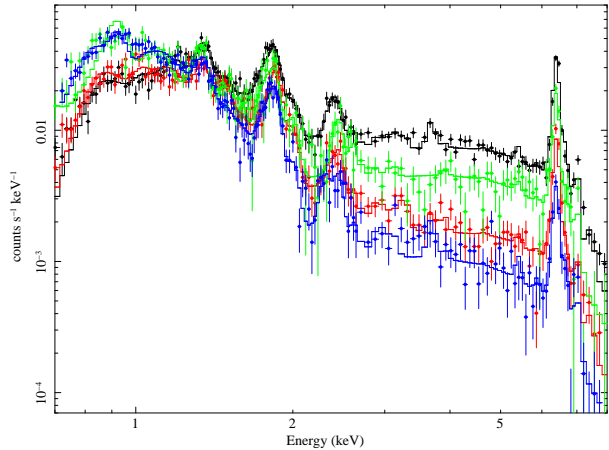}
\caption{\label{diffuse_spectra} \chandra\ ACIS spectra and models of the diffuse emission regions within the \nustar\ extraction radius: center (black), circumnuclear ellipse (red), ionization cone (blue) and everything else (green). All regions contain soft X-ray emission lines and a strong reflection continuum together with prominent neutral Fe K lines.}
\end{figure}

The ionization cone, circumnuclear region, central annulus and everything else make similar contributions to the 0.5--8~keV flux. Their spectra and best-fitting models are plotted in Fig. \ref{diffuse_spectra}. All the spectra of extended regions show a soft component with emission lines plus a hard continuum component above 2~keV and strong neutral Fe lines at 6.4 and 7~keV, characteristic of reflection off cold material. This last component is likely associated with a Compton hump extending to higher energies.    

A simple fit to the four spectra was performed using the nuclear high-resolution spectrum model described in Sec. \ref{hetg}. We allowed the normalization of the Compton-scattered component and the scattered powerlaw plus soft lines to vary independently, to account for different ratios of optically thin to optically thick reflection. We also added one thermal component to account for the hot gas (shocked and starburst related) emission. This produced reasonable fits, which were improved by allowing the normalizations of the low energy emission lines to vary as well. The resulting models are plotted as lines in Fig. \ref{diffuse_spectra}. Notice that in the Compton-scattered component (continuum plus Fe lines) only the normalization was allowed to vary, therefore the equivalent width of the Fe lines was assumed to be the same in all regions. This produced a good description of the data. 

Since the Compton-scattered and powerlaw spectral shapes were fixed, it is possible to directly compare their contributions to the flux from the model normalization. The normalization of the Compton-scattered component is 1.5\%, 3.4\%, 14.4\% and 7.8\% of the nuclear value in the ionization cone, circumnuclear region, central annulus and everything else, respectively. 

It is important to note that the Compton-scattering component in the contamination model is most likely produced by the AGN from reflection at large distances. However, the Compton scattering parameters are not well constrained by the 0.6--8.0~keV data so they will be allowed to vary in the joint fit to \chandra , \xmm\ and \nustar\ spectra. 

\subsection{Contamination model}
The aim of the previous sections was to model the spectral shape and flux of contaminating sources below 8~keV and predict their contribution above this energy in the \nustar\ band.  We found that the largest contributions above 8~keV are produced by the Compton-scattered flux from diffuse emission regions and, to a lesser extent, the thermal emission from the hottest gas in CGX2 and the powerlaw spectra from X-ray binaries. We collected these components into a simplified model for the contamination, composed of a thermal plasma component to model all the thermal emission of the diffuse regions, all the thermal components of CGX2 fixed, one Compton-scattered continuum plus Fe lines to model an average of all the off-nuclear neutral reflection, one absorbed powerlaw with $\Gamma= 1.8$ for all the point-source emission (with fixed normalization) and another soft powerlaw for the AGN scattered emission.

\begin{figure}
\psfig{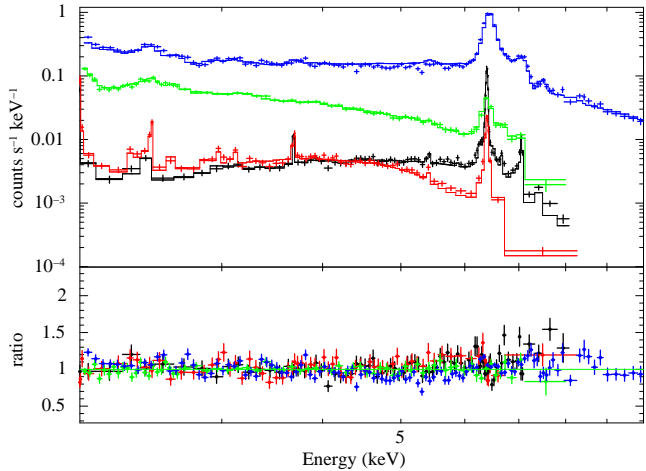}
\caption{ Joint fit of the low energy data: \chandra\ \heg\ (black) and \meg\ (red) fitted with the nuclear model, ACIS  (green) represents the contaminating sources in the galaxy and \xmm\ pn (blue), is modeled as the sum of all the nuclear and contamination models. The normalization of the variable CGX1 was allowed to vary between the ACIS contamination spectrum and the pn full spectrum and line energies were shifted by 40 eV in the pn spectrum to account for calibration differences. All other model parameters are the same for all data sets. 
\label{low_energy_joint_fit}}
\end{figure}

{ The sum of the nuclear component fit to the \hetg\ data and the contamination fit with ACIS data should account for the total flux within the \nustar\ aperture. The \xmm\ pn spectrum from 2013 extracted from the same region was fitted with the sum of both components with all emission lines above 6~keV in the model shifted up by 40~eV, including the Compton-scattered Fe lines modeled with \mytorus\ in the nuclear and contamination components. The cross-calibration between \xmm\ pn and \chandra\ ACIS has been calculated by \citet{nevalainen10} among others. These authors find a flux ratio between pn and ACIS of 0.81--1 for galaxy cluster measurements in the 2--7 keV, where the range in cross-calibration factors is much larger than the statistical error on each fit. We allow for overall normalization differences in a similar range. 
Freezing all the parameters except an overall normalization factor we model the pn spectrum as the sum of the contamination and nuclear spectral models. Since our contamination spectrum has at least one variable source and the spectra are not simultaneous, we further allow the normalization of CGX1 to vary in the pn spectrum. CGX1 is responsible for approximately a constant fraction of the continuum contamination flux in this energy band so a varying XRB flux can emulate a different cross-calibration except in the regions where the emission lines dominate the flux. Considering these two free normalization factors for the pn spectrum only, we fit the three data sets simultaneously, allowing only the normalization of the Compton-scattered component to vary for all data sets together.  This joint fit produces \redchi2 =1.13 for 1069 dof, in the 2--8 keV (2--10 keV for pn) and is shown in Fig. \ref{low_energy_joint_fit}. Inspection of the residuals shows a small but systematic difference between the \chandra\ and pn spectra, where the pn flux drops steadily with respect to the other instruments by about 10 \% throughout the energy band. Similar trends have been found at lower energies by \citet{nevalainen10}. This difference between instruments produces most of the residuals. The best-fitting cross-calibration coefficient is $1.02\pm 0.01$ and the normalization of the CGX1 powerlaw is $(1.65 \pm 0.5) \times 10^{-4}$ \norm\ at 1 keV, which represents a drop of about a factor of two of this component in the pn data compared to the ACIS spectrum. In the following fits, we will freeze the pn CGX1 normalization and cross-calibration coefficient to these best-fitting values.}

\section{Long term spectral evolution of the brightest sources in Circinus: \xmm\  observations.}
\label{xmm}
The \xmm\ observations of Circinus are separated by 12 yr with the second observation taken simultaneously with one of the \nustar\ exposures. We use these data to investigate possible long-term variations in the X-ray flux and spectrum and use the simultaneous observations to estimate the cross-calibration offset between the observatories. 

\begin{figure}
\psfig{file=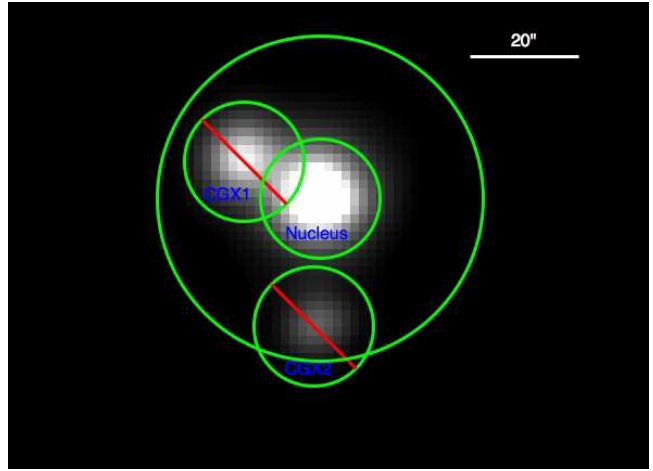,width=8.5cm,angle=0}
\caption{ \xmm\ pn image of Circinus. The central green circle 11$\arcsec$ in radius marks the extraction region for the nuclear spectrum while the background were taken from the larger annulus shown, after removing the regions shown around CGX1 and CGX2. Similar source and background regions were used for CGX1 and CGX2 spectra, centering the annuli on the source of interest and removing the other two sources from the background area. For CGX1 the extraction radius was changed to 7.5\arcsec\ to reduce the contamination from the nucleus. 
\label{xmm_image}}
\end{figure}

\begin{figure}
\psfig{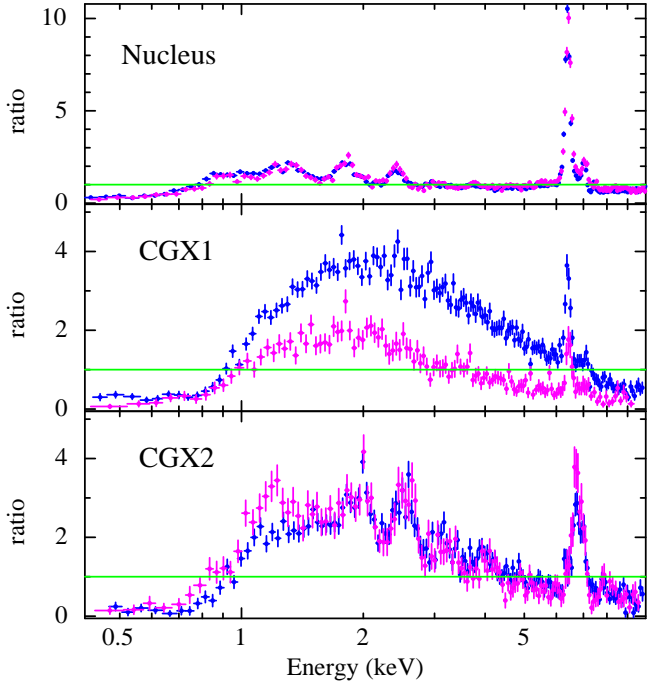}
\caption{Comparison of 2001 (blue) and 2013 (magenta) spectra of the nuclear region, the XRB CGX1 and the supernova remnant CGX2 using \xmm\ pn data. For each object, both epochs have been fitted with the same powerlaw, so that the ratios plotted highlight the evolution of each spectrum. The strongest evolution is shown by CGX1, where the flux dropped by a factor of 2. The nuclear flux and spectral shape has remained essentially constant. CGX2 shows a small drop in gas temperature and a very small change in flux. 
\label{xmm_spectra}}
\end{figure}

The lower spatial resolution of \xmm\ compared to \chandra\ allows us to distinguish only three separate sources within the \nustar\ aperture: the nuclear region and the point sources CGX1 (XRB) and CGX2 (supernova remnant). We extracted spectra for these three regions using 11\arcsec\ aperture radii for the nucleus and CGX2, and a 7\farcs5 radius aperture for CGX1 to reduce the strong contamination of nuclear flux in this region. The background region for the nuclear spectrum is shown in Fig. \ref{xmm_image} and similar regions were selected around the other two sources. The resulting spectra for 2001 (blue) and 2013 (magenta) are shown in Fig. \ref{xmm_spectra}. 

As expected, the nuclear component displays the smallest amount of variability. This spectrum was fitted with \mytorus\  scattered plus line emission, two components of hot gas and additional broad Gaussian lines, similar to the model used for the \heg /\meg\ nuclear spectrum described above. Allowing only an overall multiplicative factor to vary between both epochs, the best-fitting factor equals $1.02\pm0.008$. This error range only represents the statistical error and does not incorporate small differences in response of the instrument over the time period nor the different source positions on the CCD. The measured 0.5--10~keV fluxes were $1.19\times 10^{-11}$ \flux\ in 2001 and $1.22\times 10^{-11}$ \flux\ in 2013. The largest discrepancy between the epochs appears around the strong emission lines, where a small shift in the line energies is evident. We estimated the line centers by fitting the 5--8~keV region of both spectra with a simpler model consisting of a powerlaw with an edge at 7.1~keV and three Gaussian lines, originally at 6.4, 7.0 and 7.4~keV in the rest frame, corresponding to neutral  Fe K$\alpha$ and K$\beta$ and Ni K$\alpha$, respectively. 

The 2001 spectrum produced line energies of 6.400$\pm$0.001, 7.054$\pm$0.030 and 7.460$\pm$0.010~keV, consistent with the expected values for these lines. The 2013 spectrum, on the other hand, produced consistently higher energies for the lines, at 6.434$\pm$0.002, 7.145$\pm$0.013 and 7.490 $\pm$0.020~keV. This is an instrumental effect produced by the energy calibration at off-axis positions on chip 4 of the pn camera (M. Stuhlinger, private communication).  The MOS spectra extracted for the same observation have the Fe K$\alpha$ and K$\beta$ lines at the correct rest-frame energies, confirming that the pn lines are shifted to higher energies by calibration effects. The shift is small but important for the joint spectral fits described below; given the strength of these lines and the high signal-to-noise ratio of the \xmm\ data, this discrepancy has a large effect on the fit statistics. Therefore, in the joint fit we shift the model lines by 40~eV when considering the pn spectrum, while keeping all other parameters tied between observatories.

The largest variation between epochs is for the XRB CGX1 with an overall decrease in 0.5--10~keV flux by a factor of 2.3, from $4.14\times 10^{-12}$ \flux\ to $1.79\times 10^{-12}$ \flux\ . We fitted this spectrum with a powerlaw of fixed $\Gamma=1.8$ under a cold absorber modeled with {\tt phabs} and two Gaussian lines at 6.4 and 7~keV. The powerlaw normalization drops by a factor of 2.7 from $1.106\pm0.019 \times 10^{-3}$ in 2001 to $4.116\pm0.018\times 10^{-4}$ \norm\ at 1~keV  in 2013. A small change in column density is required by the fit,  from $1.1 \times 10^{22}$ \cm\ to $7.8  \times 10^{21}$ \cm . As seen in the ACIS spectrum of CGX1, the XRB itself does not produce emission lines, so the Fe K$\alpha$ and K$\beta$ lines observed in its pn spectrum must be contaminated by the nucleus and diffuse emission. Any contamination from the reflection continuum that accompanies these lines is incorporated into the powerlaw component of the fitted model, so the actual flux change of CGX1 alone might be larger than measured. 

The supernova remnant CGX2 evolved to lower temperatures. When modeled by three hot gas components under cold absorption using MEKAL and {\tt phabs} models as above, the temperatures change from 10.8$\pm$0.5, 1.2$\pm$0.1 and 0.12$\pm$0.03~keV in 2001 to 8.3$\pm$0.5, 0.9$\pm$0.5 and 0.13$\pm$0.01~keV in 2013. The integrated 0.5--10~keV flux changed by 3 percent, from $1.51\times 10^{-12}$ \flux\ to $1.56\times 10^{-12}$ \flux .

The analysis above shows that we can expect changes in the contaminating sources within the \nustar\ aperture at different epochs. The observations performed with \nustar\ and \chandra\ are closer in time to the second \xmm\ observation, so we will use this last \xmm\ observation to make the joint spectral fits. We bear in mind, however, that the energy calibration of the 2013 pn spectrum is shifted to higher energies by about 40~eV at 6~keV and that the normalization of CGX1's contribution to the contamination must be allowed to vary between observations.

\section{Broad-band spectral fitting}
\label{fits}
Having modeled the nuclear and contamination spectra in the 2--8~keV band, we combine these components to fit the 2--79~keV band. We use the models fit to the nuclear \hetg\ spectra and the contamination model fit to the ACIS spectrum. These two models together should reproduce the \xmm\ pn and \nustar\ spectra extracted from the entire 100\arcsec\ radius circular region centered on the AGN.  For the joint fit we use the \chandra\ spectra together with the 2013 \xmm\ pn and the on-axis 2013 \nustar\ data observation. These last two instruments cover the same nuclear and galactic sources with overlapping energy ranges. { We will also use the \swift\ BAT spectrum above 20 keV, since this spectrum is consistent with the \nustar\ data and extends to higher energies.}

All the components that are well constrained by the spectral fitting below 8~keV, including hot-gas emission and point-source powerlaws together with their cold absorbers, remain fixed in the fitting below. The remaining components correspond to scattered powerlaw, Compton-scattered and transmitted AGN continuum. These are not well constrained by the low energy spectra and have large contributions to the \nustar\ band, so they are allowed to vary in the joint fit. 

The normalization of the  CGX1 powerlaw is expected to vary from one epoch to another, so this parameter is different for the \xmm , \chandra\ and \nustar\ spectra. The normalization is fixed for the ACIS spectrum from which it was originally fit, the \emph{difference} between \xmm\ off-axis and \nustar\ on-axis is fixed as described in Sec. \ref{xmm} and the difference between the  \xmm\  and ACIS observations is fit from the joint model. This component does not  enter in the grating spectra.  The joint nuclear and contamination models slightly over-predict the Fe lines in the \xmm\ and \nustar\ spectra since the nuclear and contamination models account for some of these photons twice. Therefore, the narrow Gaussians introduced at 6.4 and 7~keV in Sec. \ref{hetg} to solve this issue in the nuclear spectrum will only contribute to the \hetg\ spectra and not to the other data sets.

For the Compton-scattered components we use different models to account for the possible different geometries of the scattering material. Below, we describe the joint fits using either \pexmon\ \citep{pexrav}, \mytorus\ \citep{ yaqoob12} or  \torus\   \citep{brightman11} to reproduce the nuclear and galactic Compton-scattered components.

The nuclear spectrum is modeled as {\tt phabs(\nh =$10^{22})\times $\{lines+AGN\}} and the contamination spectrum  as {\tt phabs(\nh =$1.16\times10^{22})\times$ \{apec +phabs(\nh =$2.2\times10^{22}$)[mekal+mekal+gsmooth(0.065 keV)$\times $\mekal ]+lines+AGN\}.}  The AGN components include scattered and Compton-scattered powerlaws, fluorescent line emission and, in the nuclear spectrum only, a transmitted continuum --- the parameters of these are summarized in Tables \ref{table_pexmon} for the \pexmon\ Compton scattering models and in Table \ref{table_mytorus} for the \mytorus /\torus\ Compton scattering models. The rest of the parameters were fitted in the previous sections and frozen in the joint fits. These are the emission line parameters, summarized in Table \ref{table_frozen_lines} and the thermal emission components summarized in Table \ref{table_frozen_continuum}.  

\subsection{Compton scattering modeled with \pexmon}
\label{pexmon}
The following analysis uses the \pexmon\ model to represent the Compton-scattered components. This model assumes that the scattering structure has a slab geometry and infinite optical depth. A directly transmitted AGN component is also allowed in the fit to the nuclear spectrum and is modeled using the zero-order \mytorus\ model, which is more accurate than a simple absorbed powerlaw model in the Compton thick regime. The Ni edge is not included in \pexmon\ so we introduced it with the model {\tt zedge} at the systemic redshift of Circinus. The depth of this feature with $\tau=0.1$ was fixed based on the measured Ni K$\alpha$ flux.   

To begin we assume that the nuclear obscurer has the same optical depth as the scatterer so that no transmitted flux reaches the observer and therefore the zero-order \mytorus\ model normalization is fixed at zero. This is the reflection dominated case. The variable \pexmon\ parameters are: the intrinsic AGN powerlaw slope $\Gamma$, the inclination angle of the scattering slab and the normalization of this component. The reflection fraction is fixed to -1, i.e. only reflected emission is considered and the reflector is assumed to cover half the sky of the primary source. The same model setup is used for the nuclear and galactic Compton scatterers. The parameters of the nuclear and galactic Compton-scatterers vary independently except for the powerlaw slope $\Gamma$ which is free to vary but tied between both these components and the scattered powerlaws. 

\begin{figure}
\psfig{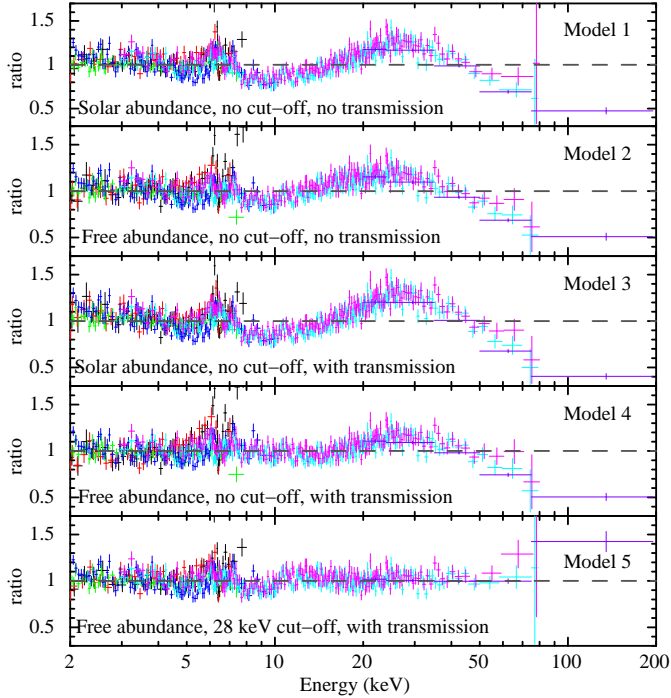}
\caption{Residuals of all the data sets to the models using \pexmon\ for the Compton-scattered components, as described in the text in Sec. \ref{pexmon}. The HEG (black) and MEG (red) nuclear data are only fit with the nuclear model, the ACIS spectrum (green) is only fit with the contamination model and the \xmm\ (blue), \nustar\ A and B (magenta, cyan) and \swift\ BAT (purple) spectra are fit with the sum of both components. The only configuration capable of reproducing the high energy curvature involves a strong transmitted powerlaw, shown in the bottom panel. This model is plotted in Fig. \ref{model_pexmon}.
\label{rat_pexmon}}
\end{figure}

We allow only $\Gamma$, the inclination angle and normalization of both \pexmon\ scatterers and the normalization of both scattered powerlaw components to vary, and keep the abundances fixed at the solar value and the cut-off energy fixed at 1000 keV. This approach gives a \redchi2 = 2.03 for 2639 dof and the residuals show a clear peak at 30--40 keV. This model is not able to reproduce the curvature of the high energy spectrum, its residuals are shown in the top panel in Fig. \ref{rat_pexmon} (model 1). 

Allowing the abundance and Fe abundance of both Compton scatterers to vary produces a better fit with \redchi2 = 1.64 for 2635 dof. 
The spectral bump at 30--40 keV is better reproduced although still apparent in the residuals, as shown in the second panel in Fig. \ref{rat_pexmon} (model 2). The fitted overall and Fe abundances for the nuclear Compton scatterer are $Z=0.83 \pm 0.05 Z_{\sun}$ and $ Z_{Fe}=1.73 \pm 0.07 Z_{\sun,Fe} $, respectively, while for the contamination component they are $Z=6.29 \pm 0.75 Z_{\sun}$ and $Z_{Fe}=0.18 \pm 0.06 Z_{\sun,Fe}$. This combination of sub- and super-solar abundances makes the Compton scattering peaks appear at different energies, below 20 keV for the nuclear component and above 20 keV for the contamination, producing  an overall broader Compton hump. This fit, however, is still unsatisfactory. 

Keeping the abundances fixed at solar and freeing the cut-off energy of the intrinsic powerlaw also gives a poor fit with \redchi2 = 1.91 for 2638 dof. In this case, the cut-off energy is tied between the nuclear and galactic Compton scatterers, so both components have similar shapes and cannot mimic a broader Compton hump. The best-fitting folding energy is $292\pm 26$~keV and the AGN continuum photon index is $\Gamma= 1.60\pm0.02$. 

As an alternative, we incorporated a transmitted AGN component in the nuclear spectrum with variable column density and normalization and repeated the procedure above with these additional free parameters. We model this component as the transmitted spectrum through an edge-on torus using the zero-order component of \mytorus\ with its inclination fixed at 90$\deg$. As above, we begin with \pexmon\ models with fixed cut-off energies well above the observed range, abundances fixed to solar values and free inclination angles. The normalizations of all scattered powerlaws, Compton-scattered components and the transmitted spectrum are allowed to vary freely. The fit is poor, with a \redchi2 = 2.14 for 2637 dof. The transmitted component normalization vanishes and therefore its \nh\ is unconstrained. The same feature as in models 1 and 2 appears in the residuals, shown in the third panel in Fig. \ref{rat_pexmon} (model 3). 
Allowing the overall and Fe abundances of both Compton scattering components to vary improves the fit as in the reflection-dominated cases, to \redchi2 = 1.44 for 2633 dof. Although this fit is significantly better it still produces a broad peak at 30 keV in the residuals, shown in the fourth panel in Fig. \ref{rat_pexmon} (model 4). 

\begin{figure}
\psfig{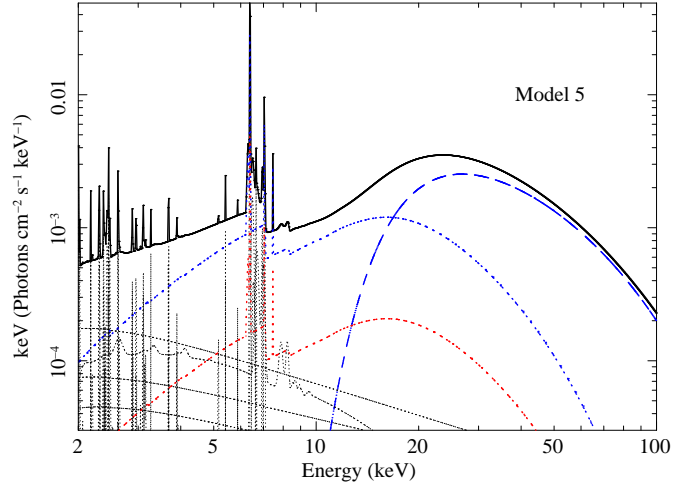}
\caption{Best-fitting model of the model group using \pexmon\ for the Compton-scattered components. The dashed blue line represents the transmitted powerlaw component (i.e. the nuclear photons that travel through the absorber unscattered). The dotted lines represent the nuclear (blue) and contamination (red) Compton-scattered components. In this model, the transmitted power dominates the total spectrum above 15-20 keV and produces 93\% of the energy in the 30--79 keV band. 
\label{model_pexmon}}
\end{figure}

We again return to solar abundances and assume a cut-off powerlaw for the intrinsic AGN component, tying the folding energy in \pexmon\ to $E_c$ of the cut-off powerlaw that is multiplied by the zero-order \mytorus\ component. This setup gives a much better fit with \redchi2 = 1.15 for 2636 dof. This model reproduces well the curvature of the \nustar\ spectra, although it under-predicts the higher energy bins covered by \swift\ BAT. The cut-off energy is fitted at $24.1 \pm0.8$ keV, the nuclear \nh =$(6.44     \pm  0.06)  \times 10^{24}$ \cm , and the photon index of the AGN powerlaw and all its scattered components is $\Gamma=1.62\pm 0.03$, the inclination angles of both \pexmon\ models converged to 0. The normalization of the nuclear (unabsorbed) powerlaw is $10.6\pm 0.8$ \norm\ at 1 keV while that of the nuclear \pexmon\ model is $0.019\pm0.001$ in the same units.
 In this model, the transmitted powerlaw component dominates the total spectrum above 20 keV. The (unabsorbed) 2--10 keV flux of the nuclear powerlaw is  $3.9 \times 10^{-8}$ \flux . The residuals to this model are shown in the last panel in Fig. \ref{rat_pexmon} (model 5) and the model itself is shown in Fig. \ref{model_pexmon}. In this model, the transmitted power dominates the total spectrum above 15--20 keV and produces 93\% of the energy in the 30--79 keV band. The best-fitting parameters to the models shown in Fig. \ref{rat_pexmon} are summarized in Table \ref{table_pexmon}.

\begin{table*}
\begin{tabular}{lllllll}
\multicolumn{6}{l}{\scriptsize Joint fits: AGN components modeled with \pexmon }\\
\hline
Component&Parameter& model 1& model 2&model 3&model 4&model 5\\
\hline
Nuclear\\
\hline
AGN&$\Gamma$&$1.675\pm0.004$& $1.82\pm0.02$&$1.74\pm 0.01$&$2.5\pm0.02$&$1.62\pm$0.03\\
continuum&$E_c$ &1000.0&1000.0&1000.0&1000.0&$24.1\pm0.6$\\
\pexmon &$Z$&1.0&$0.83\pm0.05$&1&$2.08\pm0.09$&1\\
&$Z_{\rm Fe}$ &1.0&$1.73\pm0.07$&1&$2.29\pm0.07$&1\\
&Incl. &$1.0^{+15.9}_{-1.0}$&$17.9^{+0.5}_{-6.5}$&$1.0^{+14.8}_{-1.0}$&$18.2^{+3.1}_{-5.5}$&$1.0^{+7.8}_{-1.0}$\\
&A$_P$&$1.86\pm0.024\times 10^{-2}$&$0.023\pm0.001$&$2.2\pm0.03\times 10^{-2}$&$0.140\pm0.008$&$0.019\pm0.001$\\
\mytorus &\nh & --&--&$10\pm10$&$3.82\pm 0.05$&$6.44\pm0.06$\\
&A$_Z$&0&0&$0\pm0.015$&$4.4\pm0.5$&$10.6\pm0.8$\\
Scattered pl&A&$1.4\pm 0.1\times 10^{-4}$ &$1.5\pm 0.3\times 10^{-4}$ & $1.5\pm 0.5\times 10 ^{-4}$& $0\pm 3.6 \times 10 ^{-5}$ & $1.5\pm 0.3\times 10 ^{-4}$\\
\hline
Contamination\\
\hline
\pexmon & $Z$&1.0&$6.28\pm0.74$&1&$20\pm2$&1\\
&$Z_{\rm Fe}$ &1.0&$0.18\pm0.06$&1&$0.12\pm0.02$&1\\
&Incl. &$1.0^{+19.0}_{-1.0}$&$1.0^{+18.0}_{-1.0}$&$1.0^{+18.1}_{-1.0}$&$18.1^{+3.2}_{-18.1}$&$1.0^{+12.4}_{-1.0}$\\
&A$_P$&$3.3\pm0.2\times 10^{-3}$&$2.1\pm0.3\times 10^{-2}$&$3.8\pm0.14\times 10^{-3}$&$0.215\pm0.003$& $3.4\pm0.2\times 10^{-3}$\\
Scattered pl&A&$9.8\pm 1.8\times 10^{-5}$ &$9.7\pm 2.3\times 10^{-5}$ & $7.5\pm1.9\times 10 ^{-5}$& $1.1\pm 0.5\times 10 ^{-4}$ & $9.5\pm 1.6\times 10 ^{-5}$\\
\hline
\redchi2 &&2.03 &1.64&2.14&1.44&1.15\\
\hline
\end{tabular}
\caption{Parameters of the best-fitting models that use \pexmon\ for the Compton scattering component. The photon index and cut-off energy are tied between the nuclear and contamination. The reflection fractions are fixed at -1. Units of \nh\ are $10^{24}$~\cm , angles are in degrees and energies in keV. The normalizations of the \pexmon\ components are denoted by A$_P$, of the transmitted nuclear component A$_Z$ and of the scattered powerlaw $A$ and are all in units of \norm\ at 1 keV. The transmitted component is modeled with the zero-order component of \mytorus\ for an edge-on torus (inclination=$90\deg$). The residuals to these models appear in Fig. \ref{rat_pexmon}.\label{table_pexmon} }
\end{table*}

In summary, modeling the Compton-scattered components with \pexmon\ only produces a good fit when the transmitted powerlaw is included and the intrinsic AGN powerlaw has a cut-off within the observed energy range. In this scenario, the optical depth of the obscurer and scatterer are necessarily different since \pexmon\ only allows an infinite optical depth. The cut-off energy is surprisingly low, at 28 keV and this under-predicts the highest energy bins. In this model, the AGN powerlaw photons that have traversed the obscurer un-scattered dominate the total (nuclear+contamination) spectrum above 20 keV. In the majority of these fits, including the best-fitting, transmission-dominated model, the AGN powerlaw was flat with a photon index  $\Gamma < 1.8$.

\subsection{Compton scattering modeled with \mytorus}
\label{mytorus}
 We expect that the Compton-scattered component should be modeled better with a torus model that takes into account the toroidal geometry of the scatterer, modeling the transmitted and Compton-scattered components consistently. We thus replace the \pexmon\ model with the \mytorus\ Compton-scattered component.  
 
 We consider first a uniform torus, so that the transmitted and scattered \mytorus\ components will have the same column densities and inclination angles and are illuminated by the same powerlaw continuum. As in the previous section, we also tie the photon index of the scattered and Compton-scattered components in the nuclear and contamination spectra. As a first step we consider an incident powerlaw with no cut-off in the observable range and solar metallicities. The Ni edge is not included in the \mytorus\ scattered component so we again introduced it with the model {\tt zedge} at the systemic redshift. 

\begin{figure}
\psfig{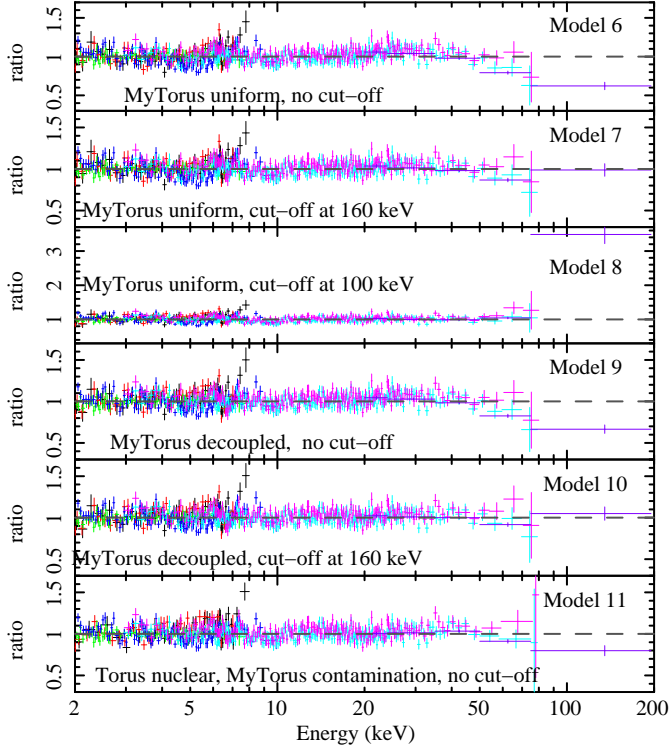}
\caption{Residuals of all the data sets to the models using \mytorus\ for the Compton-scattered components, as described in the text. The last panel uses \torus\ instead of \mytorus\ for the nuclear transmitted and Compton-scattered components together. These models reproduce well the curvature around 30 keV using solar metallicities and little or no transmitted components, consistent with the parameters of the scatterer. A cut-off in the powerlaw is required above the \nustar\ spectral range to reproduce the \swift\ BAT data points. {  Note that the cut-off energies refer to a termination energy and not an exponential roll-over of the powerlaw flux.} The models corresponding to panels 2, 5 and 6 are shown in Fig. \ref{model_mytorus}. Color are as described in Fig. \ref{rat_pexmon}.
\label{rat_mytorus}}
\end{figure}

Allowing the photon index of the AGN continuum to vary, along with the inclination angle, \nh\ and normalization of both Compton scatterers, and the normalization of the scattered powerlaws produced a fit with \redchi2 =1.195 for 2637 dof. This quality is similar to the best \pexmon\ model described above, the residuals are shown in the top panel in Fig. \ref{rat_mytorus} (model 6). {  Note, however, that the current setup is slightly more restrictive in the sense that the transmitted and Compton-scattered components are forced to have the same normalization and obscuration parameters, we allow no cut-off in the powerlaw and the metallicity is fixed at the solar value.}
nuclear=10
 
 We now relax the assumption that the nuclear obscurer and scatterer are identical by allowing the \nh\ of the nuclear transmitted and Compton-scattered components to be different. This could be the case in a patchy torus, where the line-of-sight \nh , i.e. in the transmitted component, differs from the global average \nh\, which shapes the scattered component. This results in only a small improvement in the quality of the fit,  \redchi2 =1.182 for 2636 dof and the residuals show the model still over-predicts the data above 80 keV, while the Compton hump around 30--40 keV is well reproduced. 
   
Returning to the uniform torus model but allowing the normalization of the transmitted powerlaw to vary freely with respect to the nuclear Compton scatterer represents a uniform torus with a variable covering fraction. This setup produces a similar fit to the first case, \redchi2 =1.187 for 2636 dof. In this fit the normalization of the transmitted powerlaw drops to zero and reduces somewhat the \nh\ of the nuclear scattered component from $10^{25}$ to $(8\pm 0.9)\times 10^{24}$~\cm . The most noticeable residuals are, as in all cases in this subsection, an overprediction of flux at high energies.

To deal with the high-energy slope we wish to introduce a cut-off to the AGN powerlaw. \mytorus, however, does not allow a cut-off energy as a free parameter in the Compton-scattered component but instead provides several tables calculated for different '{termination energies}'.  We replaced the powerlaw multiplying the zero-order \mytorus\ component by a cut-off powerlaw and fixed its cut-off energy to that of the {  termination energy of the} Compton-scattered table used for each fit. {  The termination energy implies an abrupt drop in the flux rather than the typical exponential rollover implied by a cut-off. Therefore, the termination energy does not represent a cut-off energy in the usual sense and can only be loosely interpreted as an upper limit of where the rollover cut-off should be. The difference in spectral shapes in the scattered components begins close to the termination energy and, as all termination energies we will use are at or above 100 keV, where the spectral resolution is poor the precise spectral shape has little impact on the fit. The main feature we explore is whether the flux at the highest \swift\ BAT energy bins is over- or under-predicted by each model.} Using a cut-off/termination energy of 160 keV results in a \redchi2 = 1.11 for 2637 dof with the corresponding residuals shown in the second panel in Fig. \ref{rat_mytorus} (model 7).  To demonstrate that 160 keV gives the best results, we refit the model using 100 keV and 200 keV cut-offs/termination energies. Lowering to 100 keV produces a worse fit, \redchi2 = 1.22 for 2367 dof, where the highest energy bins are under-predicted, as shown in the third panel in Fig. \ref{rat_mytorus} (model 8).  Increasing the energy to 200 keV also produces a slightly worse fit than $E_{\rm c}=160$ keV, with \redchi2 = 1.13 for 2637 dof.

Using the best-fitting model, with a cutoff/termination energy at 160 keV, we explore the patchy torus possibility by untying the \nh\ of the nuclear Compton scatterer from that of the transmitted component.  The fit is not much better, with \redchi2  = 1.108, and both values of \nh\ remain consistent with $10^{25}$~\cm . 
Returning to the uniform torus, but allowing for different covering fractions of the nuclear scatterer by untying the normalizations of the transmitted and nuclear Compton-scattered components, produces \redchi2 = 1.108 with a vanishing transmitted component. 

\begin{figure}
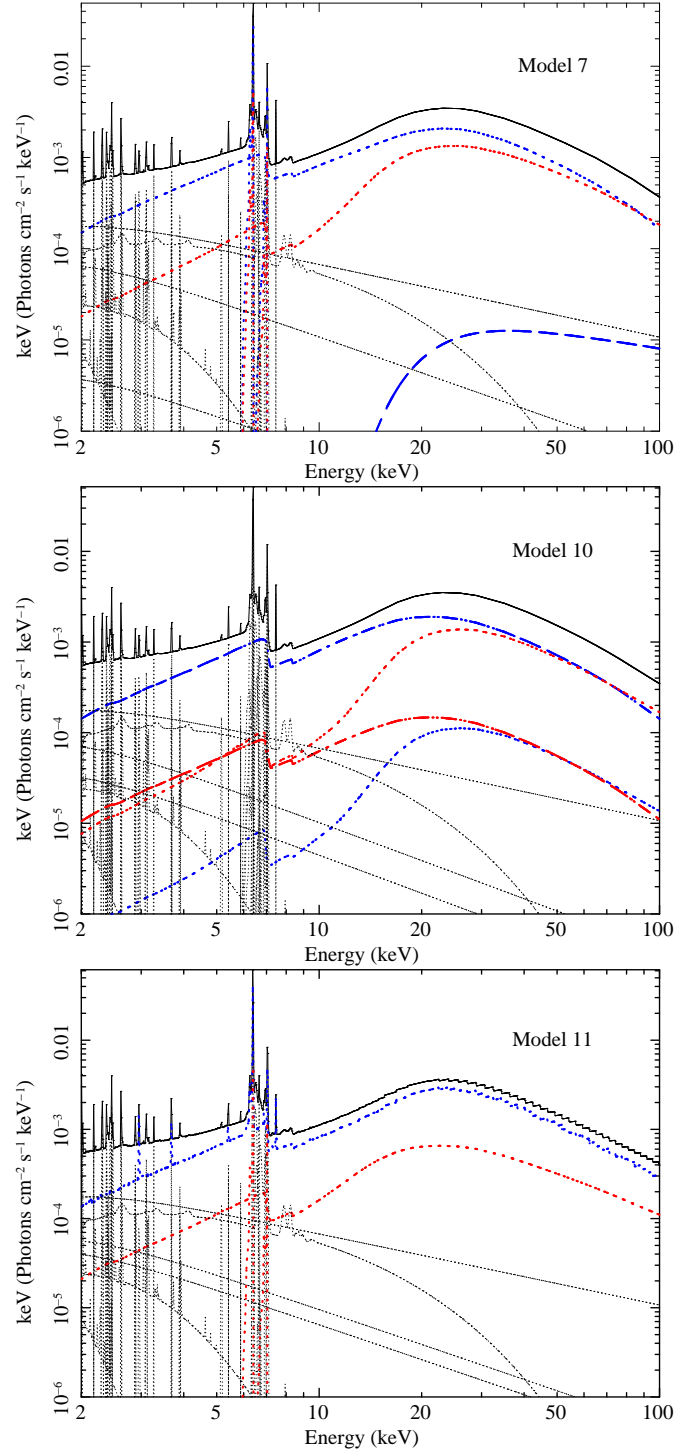

\psfig{file=model_AGN_galaxy_mytors_withBAT_model4.ps,width=9.0cm,angle=270}
\psfig{file=model_AGN_galaxy_mytorus_withBAT_model12.ps,width=9.0cm,angle=270}
\psfig{file=model_AGN_galaxy_torus_withBAT_modelo2.ps,width=9.0cm,angle=270}
\caption{Best-fitting models of the model group using \mytorus\ and \torus\ for the Compton-scattered components. Blue lines represent nuclear components and red lines contamination components. Top panel: uniform torus modeled with coupled \mytorus\ transmitted (dashed lines) and Compton-scattered (dotted lines) components. The contamination Compton scatterer  in red dashed lines becomes comparable to the nucleus at high energies. The transmitted component is negligible. Middle panel: Decoupled \mytorus\ component having forward scattered (dotted lines) and reflected (dot-dashed lines) in both nuclear and contamination spectra. As above, the nuclear and contamination scatterer components become comparable at about 30 keV and the transmitted component is negligible. Bottom panel: similar to the top panel but using \torus\ instead of \mytorus\ for the nuclear Compton-scattered plus transmitted spectra. Given the fitted torus parameters, the transmitted component (incorporated in the blue dashed line) should be a small fraction of the nuclear spectrum. The residuals of the spectra to these three models are shown in Fig.\ref{rat_mytorus} in panels 2, 5 and 6. The models are discussed in Secs. \ref{mytorus} and \ref{torus}.
\label{model_mytorus}}
\end{figure}

\subsubsection{Decoupled \mytorus\ implementation}

We also allow for a patchy torus where some reflection off the front of background clouds bypasses  the clouds towards the line of sight, as suggested in \citet{yaqoob12}. This scenario is implemented in \mytorus\ by using two Compton scatterers, one edge-on and one face-on, where the corresponding normalizations, $A_{90}$ and $A_{00}$ vary independently. Each Compton scatterer has its associated Fe K$\alpha$ and K$\beta$ lines. We use this setup for the Compton-scattered components of the nuclear and the contamination spectra. At first, we tie together the \nh\ of all four Compton-scattered components. Allowing all the Compton-scattered, scattered and transmitted powerlaw normalizations to vary independently produces a good fit with  \redchi2 = 1.106 for 2367 dof and a vanishing transmitted component. The residuals of this model are shown in the fourth panel in Fig. \ref{rat_mytorus} (model 9). 

Untying \nh\ for the nuclear and contamination spectra produces \redchi2 =1.102 for 2366 dof, a vanishing nuclear $A_{90}$ and small contamination $A_{00}$. It does not make sense to decouple the \nh\ of the forward scattering and back reflection components of the nuclear component, since they refer to the same distribution of clouds; the same reasoning applies to the Compton-scattered components in the contamination model. It can happen, however, that the line-of-sight \nh\ is different from the global average. Allowing \nh\ to vary independently for the transmitted and for the Compton scatterers in the nucleus does not produce a better fit and, in this case too, the transmitted component normalization vanishes. 

 All of these fits reproduce well the curvature of the nuclear spectrum but over-predict the highest energy \swift\ BAT spectral bins as in the case of coupled \mytorus\ models. To solve this issue we repeat our procedure with the cut-off powerlaws and termination energies. With a cut-off/termination energy of 160 keV we obtain a \redchi2 = 1.056 for 2637 dof. These residuals are shown in the fifth panel in Fig. \ref{rat_mytorus}  (model 10) and the model is shown in the middle panel in Fig.\ref{model_mytorus}.   Fitting the nuclear and contamination \nh\ values separately does not improve the fit.

\subsection{Compton scattering modeled with \torus}
\label{torus}
As the geometry of the torus is not known, we also tested the applicability of the \torus\ model \citep{brightman11}, which assumes a spherical obscurer with conical polar openings of a variable opening angle, instead of the toroidal structure with fixed opening angle used in \mytorus . {  This model combines the Compton scattered and transmitted components, which are therefore always modeled consistently. It is not possible within this model to separate the photons that went through the torus un-scattered from those that suffered one or more scatterings, such that a fraction of the \torus\ flux will correspond to transmitted photons. For the high column densities measured, however, this transmitted flux is always a small fraction of the Compton scattered flux.} \torus\ cannot be used to model the Compton scattering of the contamination spectrum since this spectrum is necessarily out of the line of sight to the AGN and therefore does not contain a transmitted component. 

We model the Compton scattering in the contamination spectrum first with \pexmon\ and then with \mytorus . {  \torus\ self-consistently predicts the K$\alpha$ emission lines of all the relevant elements, so we removed Gaussian lines at 2.96, 3.69, 5.41 and 7.47 keV, corresponding to the K$\alpha$ transition of Ar, Ca, Cr and Ni, from the nuclear model to avoid double-counting them. }Keeping all the AGN-related photon indices tied, a free torus opening angle, free inclination and free normalizations of the scattered powerlaws and contamination Compton scattering component produces a  \redchi2 = 1.24 for 2636 dof. We now replace \pexmon\ in the contamination model with \mytorus . Repeating the fit above, with free \nh\ and inclination angle values for the \mytorus\ Compton-scattered component produces a better fit with \redchi2=1.125 for 2636 dof. The residuals to this model are shown in the bottom panel in Fig. \ref{rat_mytorus} (model 11) and the model is shown in the bottom panel in Fig. \ref{model_mytorus}. This model produces a better fit than \mytorus\ in the default configuration when no cut-off is allowed below 500 keV in the incident powerlaw, and it only overestimates the highest energy BAT points by tens of percent. We contend that a cut-off at similar energies as the \mytorus\ case would improve this fit further but it is not possible to introduce this feature with the current implementation. 

{  In summary, the use of either \torus\ or \mytorus\ produces a better description of the spectra than  \pexmon\ when we consider reflection-dominated models. As mentioned in the previous subsections, \pexmon\ can  produce an acceptable fit when a strong transmitted component is incorporated. In all the fits with \torus\ and \mytorus\, }the transmitted component is negligible in the observed energy band, the column densities of the nuclear absorber range from (8--10)$\times 10 ^{24}$~\cm\ and the AGN photon index is in the range  $\Gamma=2.2-2.4$. Using cut-off/termination energies above 100 keV and solar metallicities, which are inherent in the models, prove adequate. Different geometries of the obscurer/scatterer produce similar quality fits so it is not possible to discriminate between a patchy or homogeneous torus or between a toroidal or spherical structure. The best-fitting parameters of the models used in Fig. \ref{rat_mytorus} are summarized in Table \ref{table_mytorus}. {  We note that the \mytorus\ tables are only calculated for column densities up to \nh =$10^{25}$ \cm\  implying that some of the 1$\sigma$ error ranges for this parameter in the Table \ref{table_mytorus} saturate at the maximum available value and a significant upper limit cannot be obtained. The \torus model, however, is calculated for column densities up to \nh =$10^{26}$ \cm, and has a best-fitting value of $8.9\pm1.2$ such that it is completely constrained within the available range}.

\begin{table*}
\begin{tabular}{llllllll}
\multicolumn{6}{l}{\scriptsize Joint fits: AGN components modeled with \mytorus\ and \torus }\\
\hline
Component&Parameter& model 6& model 7&model 8&model 9&model 10&model 11\\
\hline
Nuclear\\
\hline
AGN &$\Gamma$&$2.34\pm0.02$&$2.31\pm0.02$&$2.19\pm 0.02$&$2.40\pm0.02$&$ 2.4\pm 0.03$&$2.31\pm 0.02$\\
continuum&$E_c$ &500&160&100&500&160&--\\
\mytorus /&\nh&$10.0\pm 1.8 $&$10\pm 2.2$&$6.6\pm 0.9$&$7.1\pm0.2$&$7.4\pm 0.2$&$8.9\pm1.2$\\
\torus&Incl. &$76\pm2.4$&$78.4\pm1.4$&$75.8\pm 2.4$&--&--&$59.2\pm 2.3$\\
&torus angle&--&--&--&--&--&$34.9\pm 2.2$\\
&A$_Z$/A$_{S}$&$1.17\pm0.36$&$1.54\pm 0.36$&$0.83\pm 0.26$&0&0&$0.66\pm 0.03$\\
&A$_{S90}$&--&--&--&$0.008\pm0.8$&$0.19\pm 0.77$&--\\
&A$_{S00}$&--&--&--&$0.44\pm0.01$&$0.43\pm0.02$&--\\
Scattered pl&A&$0\pm 5\times 10^{-5}$ &$1.2\pm 5.7\times 10^{-5}$ & $1.5\pm 0.5\times 10 ^{-4}$& $1.5\pm 0.5\times 10 ^{-4}$ & $1.1\pm 0.6\times 10 ^{-4}$&$1.3\pm 0.5\times 10 ^{-4}$\\
\hline
Contamination\\
\hline
\mytorus &\nh & $4.6\pm 0.2$&$5.0\pm 0.2$&$7.0 \pm 5.6$&7.1&7.4&$3.9\pm 0.5$\\
&Incl. (deg) &$85.1\pm 0.4$&$84.7\pm 0.8$&$80.0\pm6.3$&--&--&$79.1\pm 1.6$\\
&A$_{S}$&$0.95\pm0.14$&$0.81\pm 0.17$&$0.26\pm0.36$&--&--&$0.24\pm 0.07$\\
&A$_{S90}$&--&--&--&$2.12\pm 0.7$&$2.35\pm 0.74$&--\\
&A$_{S00}$&--&--&--&$0.036\pm 0.012$&$0.03\pm0.74$&--\\
Scattered pl & A & $2.3\pm 0.4\times 10 ^{-4}$& $2.2\pm 0.4\times 10 ^{-4}$& $1.8\pm 0.4\times 10 ^{-4}$& $2.6\pm 0.5\times 10 ^{-4}$ & $2.6\pm 0.5\times 10 ^{-4}$& $2.0\pm 0.4\times 10 ^{-4}$\\
\hline
\redchi2&&1.195 &1.11&1.12&1.106&1.06&1.125\\

\hline
\end{tabular}
\caption{Parameters of the best-fitting models that use \mytorus\ for the Compton scattering component. The photon index and cut-off/termination energy are tied between the nuclear and contamination spectra.  Units of \nh\ are $\times10^{24}$~\cm . The normalizations of the \mytorus\ Compton-scattered components are denoted by A$_S$ in the traditional implementation, and A$_{S90}$, A$_{S00}$ for the forward scattering and back reflection components in the decoupled implementation. The normalizations of the transmitted nuclear component  and  the scattered powerlaw are A$_Z$ and $A$, respectively. In the default configuration models quoted here, $A_Z$ and $A_S$ were tied. Untying these normalizations resulted in vanishing $A_Z$. All normalizations are in units of \norm\ at 1 keV, angles are in degrees and energies in keV. The residuals to these models appear in Fig. \ref{rat_mytorus}.\label{table_mytorus} }
\end{table*}

\begin{table}
\begin{tabular}{lllllll}
\multicolumn{5}{c}{\scriptsize Emission lines in the nuclear and contamination spectra}\\
\hline
Energy&sigma&Norm.&Norm.&ID \\
(keV)&(keV)&Nuclear&Cont.\\
\hline
2.005& 1.7$\times 10^{-3}$&1.08$\times 10^{-5}$&1.02$\times 10^{-5}$&Si XIV\\
2.308&4.7$\times 10^{-3}$&7.12$\times 10^{-6}$&4.35$\times 10^{-6}$&S II-X \\
2.377&3.7$\times 10^{-3}$&5.95$\times 10^{-6}$& 2.45$\times 10^{-6}$&Si XIV+ S XII\\
2.400&7.9$\times 10^{-3}$&1.97$\times 10^{-6}$ &9.96$\times 10^{-7}$&S XIV \\
2.435&9.3$\times 10^{-3}$  &9.57$\times 10^{-6}$&-- &S XV\\
2.46&2.1$\times 10^{-3}$& 8.31$\times 10^{-6}$& 7.07$\times 10^{-6}$&S XV\\
2.62&1.4$\times 10^{-3}$&4.88$\times 10^{-6}$&7.40$\times 10^{-6}$ &S XVI \\
2.88& 2.9$\times 10^{-3}$&2.17$\times 10^{-6}$&1.85$\times 10^{-6}$&S XV \\
2.96&5.7$\times 10^{-3}$& 3.30$\times 10^{-6}$& 6.18$\times 10^{-7}$&Ar II--XI \\
3.27&1.2$\times 10^{-3}$&--& 2.6$\times 10^{-6}$&Ar XVIII\\
3.90&4.8$\times 10^{-3}$& 5.30$\times 10^{-7}$& 1.07$\times 10^{-6}$&Ar XVIII\\
3.69&1.0$\times 10^{-3}$&3.00$\times 10^{-6}$&-- &Ca II--XIV\\
&&&&+ Ar XVII \\
3.686& 1.0$\times 10^{-3}$&2.65$\times 10^{-6}$&5.42$\times 10^{-7}$&Ca II--XIV\\
&&&&+ Ar XVII\\
6.66&1.0$\times 10^{-2}$& 9.97$\times 10^{-6}$&--&Fe XXV \\
6.95 & 1.0$\times 10^{-2}$ &7.30$\times 10^{-6}$&--&Fe II--XVII\\
2.18&1.5$\times 10^{-6}$& 3.00$\times 10^{-6}$& 3.69$\times 10^{-6}$&Si XIII K$\beta$\\
3.11&5.8$\times 10^{-3}$& 2.90$\times 10^{-6}$& 2.04$\times 10^{-6}$&S XVI K$\beta$\\
5.41&1.0$\times 10^{-4}$& 3.70$\times 10^{-6}$& 1.58$\times 10^{-6}$&Cr I\\
5.88&1.0$\times 10^{-3}$  &3.11$\times 10^{-6}$ &4.76$\times 10^{-7}$&Cr XXIV \\
6.50&5.0$\times 10^{-2}$& 3.30$\times 10^{-5}$& 7.81$\times 10^{-6}$& FeXXV triplet\\
7.47&1.0$\times 10^{-3}$& 1.36$\times 10^{-5}$&--&Ni I\\
\hline
\end{tabular}
\caption{Emission lines fitted to the nuclear spectrum, the first 15 from \citet{sambruna01}, where line energies were allowed to vary within the uncertainties quoted therein. Line widths were fitted to the nuclear spectrum and frozen to those values in the fit to the contamination spectra. Unless otherwise specified, lines correspond to K$\alpha$ transitions. Lines with vanishing normalizations were eliminated. All these parameters were frozen in the joint fits described in Sec. \ref{fits}. \label{table_frozen_lines} }
\end{table}

\begin{table}
\begin{tabular}{llllll}
\multicolumn{6}{c}{\scriptsize Continuum components in the contamination spectrum}\\
\hline
Model&kT &Z&z&$\Gamma$&Norm.\\
\hline
\apec &0.238&1.0&1.45$\times 10^{-3}$&--&8.66$\times 10^{-3}$\\
\mekal&0.09&0.5&7.80$\times 10^{-2}$&--&2.37\\
\mekal&1.09&7.5&-6.58$\times 10^{-3}$&--&1.98$\times 10^{-4}$\\
\mekal&9.15&7.5&-2.86$\times 10^{-3}$&--&6.33$\times 10^{-4}$\\
powerlaw&--&--&--&1.80&4.27$\times 10^{-4}$\\
\hline
\end{tabular}
\caption{Continuum components in the contamination spectrum, corresponding to hot diffuse gas, a SN remnant and the powerlaw spectrum of CGX1 and other X-ray binaries together. All these parameters were frozen in the joint fits described in Sec. \ref{fits}. \label{table_frozen_continuum}}
\end{table}

\section{Timing constraints on the high-energy spectrum}
\label{variability}
One way to differentiate between a transmission dominated scenario, as preferred by the \pexmon\ modeling, and a Compton-scattering dominated scenario, as preferred by the \mytorus\ and \torus\ models, is by their temporal behavior. In particular we note that the direct AGN X-ray continuum is highly variable, as has been observed in large samples of unobscured sources. If these AGN are intrinsically the same as obscured sources, then the primary continuum should be equally variable in obscured AGN. Therefore, at high energies where the continuum might ``punch" through the obscuring material, obscured sources should be as variable as un∫obscured ones. One example is the obscured AGN NGC 4945, where the lower obscuration towards the nucleus allows a variable powerlaw to be seen at high energies as suggested by \citet{iwasawa93} using {\sl Ginga} data and studied in detail with \nustar\ (Puccetti et al. \emph{submitted}).

We constructed power spectra of the high-energy \nustar\ lightcurves to compare with the expected power spectrum of a transmitted component. We used {\sc xselect} to extract 30--79~keV photons for the source region and for a background region of equal area on the same detector.  Light curves were then constructed by binning the photons in 100 s equally spaced bins of which only those with exposure ratios over 90\% were retained. The low orbit of \nustar\ produces 2 ks gaps in the lightcurves every 6 ks cycle. The first observation produces lightcurves covering 90~ks, while the second observation is split into three segments of 30--80~ks duration separated by two gaps of 60 and 100~ks. We computed separate power spectra for each module and for the on-axis and off-axis observations using the Mexican-hat filtering method described in \citet{rmsk}. This method can easily combine the different segments of the second observation and is largely unaffected by gaps in the lightcurves.  In this normalization, the variabiltiy power equals the variance of the filtered light curve divided by the average count-rate squared. 

\begin{figure}
\psfig{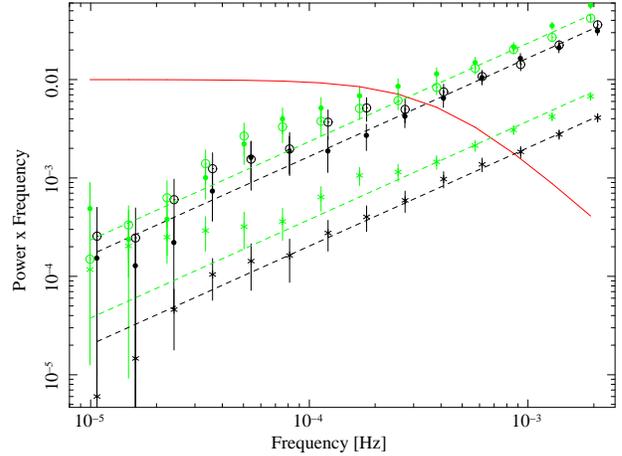}
\caption{Power spectra of the high energy (30--79~keV) \nustar\ lightcurves. The on-axis observations are plotted in black and off-axis in green, while filled circles represent FPMA source lightcurves and open symbols represent FPMB source lightcurves. The power spectra of the background regions only is plotted in stars following the same color convention. The dashed lines show the power spectra expected for Poisson noise only for each case. The solid red line represents the power spectrum of the direct continuum for an AGN of the same mass and accretion rate as Circinus. The high-energy lightcurves are consistent with simple Poisson noise and their variability is far below the expected value for a transmitted AGN continuum.  
\label{pds}}
\end{figure}

The observational Poisson noise in the lightcurves contributes variability power with a white noise (flat) spectrum. In Fig. \ref{pds} we plot power$\times$frequency so this flat spectrum appears as a linear rising trend. The normalization of the expected Poisson noise spectrum for each lightcurve was calculated from the source and background counts in each time bin and the resulting spectra are plotted as dashed lines. The on-axis observation produced slightly smaller errors and therefore lower Poisson noise than the off-axis observation. 

The power spectrum of AGN continuum emission has been measured for many type 1 Seyfert galaxies, where the 2--10~keV range is largely unobscured. The power spectra typically have a broken powerlaw shape, where the low and high frequency slopes are $-1$ and $-2$ respectively \citep[e.g.][]{markowitz03,McHardy4051,McHardyMCG,mchardynat,markowitz07}. The normalization below the break frequency is very similar between different AGN. When the power spectra are normalized by the average count rate squared, the dimensionless quantity power $\times$ frequency is typically in the range 0.01--0.1. The distinguishing feature of the power spectra of different AGN is the position of the break timescale, which depends on the black-hole mass and accretion rate. \citet{mchardynat} fit the relation
\begin{equation}
\log{T_B}=2.1 \log{M}-0.98\log{L_{\rm bol}}-2.28
\end{equation}
where $T_B$ is the break timescale in days, $M$ is the black hole mass in units of $10^6$\msun\  and $L_{\rm bol}$ is the bolometric luminosity in units of $10^{44}$~\ergs . Assuming $M=1.5$ and $L_{\rm bol}$=0.4 for Circinus, the break timescale is expected at 0.03 days and the corresponding break frequency is $4\times 10^{-4}$ Hz. A bending powerlaw model with low-frequency normalization of 0.01 and this break frequency is plotted as a solid red line in Fig. ~\ref{pds}. If the high-energy emission we observed in Circinus were mostly a transmitted powerlaw component we would expect its power spectrum to resemble this curve.  

\begin{figure}
\psfig{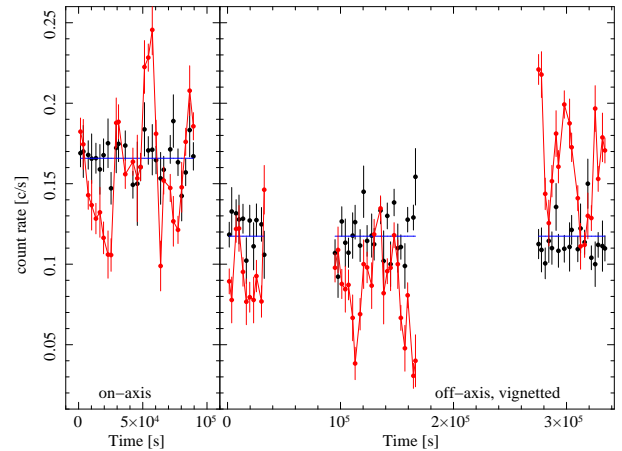}
\caption{Circinus \nustar\ lightcurves in 30--79~keV band, A and B module data combined, binned in 3 ks, in black markers with error bars. The horizontal solid blue lines represent the average count rate for each observation; the fit to this model produces $\chi ^2/dof=20/28$  for the on-axis observation (left) and 68/53 for the off-axis observations (right). The red markers and lines are a random realization of a lightcurves whose underlying power spectrum corresponds to the red line in Fig. \ref{pds}, i.e., the expected amplitude and timescales of variability for the intrinsic powerlaw emission for an AGN of the same mass and accretion rate as Circinus.   The difference in average count rate between both observations is caused by vignetting; their calibrated fluxes are the same.
\label{lcs}}
\end{figure}

The measured power spectra follow closely the Poisson noise level and are far below the intrinsic variability spectrum expected from a transmitted component. Note that symbols of equal color correspond to simultaneous observations so the difference between their filled and open circles are due to differences in the Poisson noise, background contribution or other systematic effects. For comparison we also plot the background power spectra for module B of each observation, where no intrinsic variability is expected. The off-axis background spectrum deviates most strongly from the Poisson spectrum at low frequencies. This additional variability in the background can contribute to the source spectrum as well, so similar amplitude deviations can be attributed to unaccounted variations in the background count rate.  Fig. \ref{lcs} shows the observed \nustar\ lightcurves combining FPMA and B data in black markers and the average count rate for each observation in blue solid lines. The error bars represent the Poisson noise expected in these background-subtracted bins. For a visual comparison of the variability we would expect to see in the powerlaw component, we over-plotted a random realization of a lightcurve whose underlying power spectrum follows the red model in Fig. \ref{pds}. The simulated lightcurve was sampled to match the observed 100 s binned \nustar\ lightcurve, scaled to the average count rate, and Poisson noise was added at the appropriate level. This simulated lightcurve was then binned in 3 ks bins, in the same way as the real lightcurves and is plotted in red markers and lines in Fig. \ref{lcs}. 

The minimal variability observed in the lightcurves is largely consistent with a constant flux affected by Poisson noise. This is the expected behavior if the high-energy flux indeed arises from Compton-scattered emission over large distances, i.e.\ across the torus, and not to a transmitted powerlaw component. We note, however, that the power spectra are normalized by the square of the average count rate. If the 30--79~keV flux contained both a variable transmitted component and a constant Compton-scattered component, then the normalized variance of the transmitted component would be reduced by the contribution of the Compton-scattered flux to the normalization. The difference of about a factor of 10 between the upper limit on the measured power at low frequencies and the expected power spectrum limits the maximum contribution of the transmitted flux to about a third of the total flux in this band. { Therefore, under the assumption that the intrinsic X-ray continuum of obscured and unobscured AGN behave in the same way, the spectral decomposition where the transmitted component contributes to 90\% of the flux in this band, such as the last \pexmon\ model described, can be ruled out. }

We explored longer term variability in the hard X-ray band using the \swift\ BAT 70-month data taken between 2004 and 2010. We found that the full 14--100~keV band lightcurve of a region centered on Circinus, binned in 1 month intervals, shows significant variability, but note that the telescope PSF of 17\arcmin\ allows for significant contamination of nearby sources. Splitting the data into several energy bands shows that the variability is only present in the softest BAT band. We constructed lightcurves in five energy bands 14--20, 20--24, 24--35, 35--50, 50--75 and 75--100~keV, binned in $10^7$ s bins to produce uniform signal-to-noise lightcurve points.  Each lightcurve contains 18 points. Fitting a constant value to each lightcurve produced \redchi2=5.21 for the first band and values between 0.67 and 1.51 for all other bands. This constant flux null-hypothesis can be rejected with high significance only for the 14--20~keV band, with a fractional rms =32\%. In all other bands the significance of the variability is less than 95\%.  Sec. \ref{xmm} shows that the nuclear flux within 11\arcsec\ up to 10~keV has not changed between two observations taken 12 years apart, so that the 14--20~keV variability detected by BAT probably corresponds to variations in contaminating sources within its field of view. The BAT extraction region includes the entire  Circinus galaxy, the bright and variable CGX1 and ULX5, as well as other potential sources in their vicinity. These XRBs, however, have softer spectra than the nucleus, and at high energies, above 20~keV, their contribution is negligible. We note that although the angular resolution of \nustar\ cannot separate the nucleus from CGX1, our \nustar\ extraction is not contaminated by ULX5 or other sources further away, which do pose a problem to all other hard X-ray observatories. 

Combining the \swift\ BAT lightcurves above 20~keV to improve the signal to noise ratio and fitting to a constant results in \redchi2=1.48, which rejects the hypothesis of constant flux with less than 90\% significance and has a fractional rms of only 7.7\%. The count rate in the 20--100~keV lightcurve is over 3 times higher than in the variable 14--20~keV, so the lack of significant variability is not due to larger uncertainties on the lightcurve bins. We conclude that the long term behavior of the high-energy nuclear flux is consistent with a constant value over the 70 months probed by BAT. The lack of high-energy variability in the BAT lightcurves is consistent with the Compton-scattered dominated spectrum, since a variable transmitted powerlaw would be more dominant at higher, not lower, energies. Therefore, if the 14--20~keV variability originated in the nuclear powerlaw, the higher energy bands would be even more variable.

The variability analysis rules out the scenario where the AGN continuum ``punches" through the obscurer in order to dominate the spectrum above 30~keV. The pure Compton scattering scenario is therefore preferred.

\section{Discussion and Conclusions}
\label{conclusion}

We presented the X-ray spectra of Circinus using a combination of \nustar , \chandra , \xmm\ and \swift\ data.

{ The high angular resolution \chandra\ data were used to separate the nuclear (central $6\arcsec$ in diameter) spectrum from the contaminating sources within Circinus that fall into the \nustar\ PSF. These contaminating sources were modeled individually and their spectra extrapolated into the \nustar\ band. Both the nuclear emission and the contamination have strong signatures of cold reflection off dense material, with high EW fluorescent Fe lines and hard continua indicative of a Compton hump. We combined the different instruments to produce a broadband spectrum and modeled it with a combination of the nuclear and contamination models. The Compton-scattered components were modeled using the \pexmon , \mytorus\ and \torus\ models. 

Compton scattering modeled with \pexmon\ on its own was unable to reproduce the curvature of the high-energy spectra, in particular the hump at around 30 keV. This model assumes the reflecting material is a slab of infinite optical depth, which places the Compton hump at around 20 keV. In order to shift the hump to higher energies and produce additional curvature, it either requires super-solar abundances or  a low energy cut-off in the incident powerlaw. {  This model is useful for applications requiring accretion disk reflection, where the column density is expected to be of the order of $\sim 10^{31}$~\cm\ \citep{svensson94} and a slab geometry is expected. However, its applicability is limited for other scattering media, such as the torus or galactic scale reflection nebulae, which can have significantly lower column densities and different geometries.}. The only configuration where \pexmon\ produced a good fit included a cut-off in the original powerlaw at the relatively low energy of 28 keV and a transmitted powerlaw component of much higher normalization that dominates the total spectrum above 30 keV. This model was subsequently ruled out by the lack of variability seen in the 30--79 keV band, which, according to this spectral decomposition, should correspond to the transmitted component. 

Previous fits to the X-ray spectrum of Circinus using the \pexrav\ model for the Compton-scattered component produced flat incident powerlaw slopes ($\Gamma \sim 1.3-1.7$) and low cut-off energies of about 50--80~keV \citep{matt99, dadina07}, similar to our results using \pexmon . The highest-energy \nustar\ and  \swift\ BAT spectra, however, are under-predicted by these models, mainly due to the low energy of the powerlaw cut-off. 

The models \torus\ and \mytorus\ produce Compton-scattered and transmitted components self consistently through a dense material of finite and variable column density. Both models have axisymmetric geometries, a sphere with polar conical holes in the first case and a torus in the second. We tested different configurations, considering smooth and patchy implementations of the \mytorus\ model for the nuclear and contamination components, as well as a spherical torus modeled with \torus\ for the nucleus plus \mytorus\ or \pexmon\ components for the contamination. In all the \mytorus\ configurations, a cut-off/termination energy in the intrinsic AGN continuum at about 160 keV improved the fit. \torus\ does not incorporate cut-off energy as a model parameter, but this model produced a better fit than the default configuration \mytorus\ model with no cut-off. Therefore, it is possible that \torus\ would produce the best fit if a cut-off energy were implemented in this model. Given this limitation, it is not possible to prefer one geometry over the other. 

Below we summarize the common features of the best-fitting models for all the torus geometries tested.  Both models reproduce the spectrum of Circinus accurately without resorting to super-solar metallicities. A cut-off in the AGN intrinsic powerlaw spectrum was preferred by all the different configurations tested; this cut-off/termination was found to be between 100 and 200 keV with a best fit at 160 keV, instead of the 28 keV of the \pexmon\ fit. {  We stress again that the \mytorus\ termination energy is not equivalent to a rollover in the intrinsic powerlaw, which is usually referred to as a cut-off, but rather to an abrupt decrease in intrinsic  flux.  Therefore, these energies should not be interpreted as cut-off energies in the usual sense. }The  nuclear column densities are in the range \nh =$ (6.6 - 10)\times 10^{24}$ \cm\ and photon indices of the incident AGN spectrum in the range $\Gamma=2.2-2.4$, much steeper than in the \pexmon\ fits.
The differences in \nh\  and $\Gamma$ values between models are significantly larger than the statistical errors, as can be seen in Table \ref{table_mytorus}, and thus are only meaningful when considering a particular geometric model.  }

The high line-of-sight column density only allows a very small fraction of the intrinsic AGN continuum to shine through, even at the highest energies probed. { In the best-fitting models, the transmitted component only accounts for 7\% of the 30--80 keV flux.} The variability of the high-energy (30--79~keV) emission is consistent with counting noise only so no intrinsic variability is detected in this band. By comparing the amplitude and timescales of the variability of similar but unobscured sources we conclude that the high-energy emission corresponds to Compton-scattered rather than transmitted powerlaw continuum, in agreement with the favored spectral models. 

In unobscured sources, where the powerlaw slope can be measured with less uncertainty, a slope of  $\Gamma=2.3$ is not unusual, although it is towards the high end of the distribution of measured slopes. The correlation found between  $\Gamma$ and accretion rate, however, does point to steeper slopes at higher accretion rates. For the values from the literature quoted in Sec. 1, the accretion rate in terms of Eddington luminosity in Circinus is $L/L_{\rm Edd}= 0.2$. According to \citet{shemmer06} the corresponding powerlaw slope is about $\Gamma=2.1$, and \citet{fanali13} predicts it to be $\Gamma=2.4\pm0.1$, straddling our measured values of $\Gamma=2.2-2.4$.

 The unobscured 2--10~keV powerlaw flux in {  the default configuration of \mytorus\  with termination energies of 500, 160 and 100 keV (models 6, 7, 8, respectively) as well as in the \torus\ model fit (model  11) is in the range $(1.1-2.4)\times 10^{-9}$ \flux . The decoupled \mytorus\ models with powerlaw cut-off/termination energies of 500 and 160 keV, where the normalizations of the Compton scattered and transmitted components vary freely (models 9 and 10) resulted in vanishing transmitted components such that an  intrinsic flux could not  be estimated for them.}  This range of fluxes corresponds to an isotropic 2--10 keV luminosity of $(2.3-5.1)\times 10 ^{42}$ \ergs . This equals 6--13\% of the bolometric luminosity estimated by \citet{moorwood96} from mid-IR spectroscopy, where most of the nuclear mid-IR flux is identified with reprocessed AGN emission. This $L_{2-10}/L_{\rm bol}$ ratio is within the range of X-ray to bolometric luminosity ratios  found in type 1 AGN \citep[e.g.][]{elvis86,fanali13}.  
 
 \citet{gandhi09} calculate the $L_{2-10 {\rm keV}}/L_{\rm 12\mu m}$ using nuclear IR luminosities, as opposed to the large aperture ISO values quoted above, finding an almost linear relation with 0 intercept. \citet{asmus13} measured a nuclear $12\mu$m IR flux for Circinus of $8.3\pm 1.0$ Jy, using ground based high angular resolution imaging. This nuclear IR value corresponds to a $L_{\rm 12\mu m}=4.3 \times 10^{42}$~\ergs\ which is within the range of unobscured $L_{2-10 {\rm keV}}$ luminosities quoted above, precisely as predicted by the relation of  \citet{gandhi09}. {  The transmitted component in the best-fitting \pexmon\ model, on the other hand, has a  2--10 keV flux of $4.0\times 10^{-8}$ \flux , corresponding to an unobscured luminosity of $L_{2-10 {\rm keV}}=8.4\times 10 ^{43}$ \ergs , 36 times larger than that expected from the $12\mu$ m flux. The fact that the interpretation of the high energy X-ray spectrum as transmitted rather than Compton scattered emission results in a \emph{larger} intrinsic X-ray flux might seem counter intuitive. At these high column densities, however, the suppression of the transmitted powerlaw by Compton scattering is very significant, so that the Compton scattered component dominates over the transmitted component when they are modeled self consistently. For the transmitted powerlaw to reach the observed 30--80 keV fluxes its intrinsic luminosity needs to be about 40 times larger than that of the powerlaw required to produce the same flux level as a Compton scattered component.} 
 
The total --- nuclear plus contamination--- Compton-scattered fractions are 0.43--0.65\% of the intrinsic AGN continuum in the 2--10 keV band and 5.8--7.7\% in the 2--100 keV band for the \mytorus\ models and 1.1\% and 12\% in the same energy ranges for the best-fitting \torus\  model. Note that in the last model the Compton-scattered component incorporates the transmitted flux, so these ratios are slightly overestimated. 

The galactic Compton scattering probably represents AGN continuum emission scattered by dense clouds in different locations of the galaxy. In the best-fitting model, the average spectrum of these distant scatterers corresponds to clouds of \nh $= 3.9-7.4\times 10^{24}$~\cm . These values are similar to but in most cases lower than those found for the nuclear Compton scatterer (i.e., the torus) of \nh $= 6.6-10\times10^{24}$~\cm . The galactic Compton scattering spectrum is therefore harder than the nuclear component, as can be seen in the dotted red lines in Fig. \ref{model_mytorus}. Similar large-scale reflection nebulae have been observed in the Galactic center, where strong Fe fluorescent emission tracks the position of known molecular clouds \citep{sunyaev93,koyama96,murakami01}. These lines together with their hard X-ray continua \citep{revnivtsev04} are interpreted as reflection of past AGN activity in Sgr A*. Modeling the X-ray spectrum of these reflection nebulae, \citet{ponti10}  find column densities of up to $8\times 10^{23}$~\cm , somewhat lower than our findings in Circinus. Given the large gas content of Circinus compared to our Galactic center, we can probably expect a larger covering fraction and column densities in the former. {  We also note  that since the strength of the galactic Compton scattering component should scale with the intrinsic AGN power, a similar contamination is likely also important in higher luminosity sources.  For lower spatial resolution observatories as well as for more distant AGN, contamination of the nuclear spectrum can be significant.}

The Compton-scattered component in the Circinus galaxy appears in all directions, not only towards the ionization cone, which should have a clear view of the nucleus. The lack of `shadow' regions in the diffuse gas reflection argues for a patchy or leaky torus, which allows the AGN continuum to illuminate material in all directions. This type of clumpy torus model is also favored by the shape of the IR spectra of both type 1 and type 2 AGN \citep{elitzur06}. The nuclear distribution of the Compton-scattered light from Circinus within a few arcsec from the center was recently studied by \citet{marinucci13}, who inferred a clumpy Compton-thick scatterer about 100 pc across. The implications for the shape of this Compton scattering material from large-scale reflection features will be studied further in a separate paper (Bauer et al. \emph{in prep}).

Finally, we studied the long--term variability of sources in the Circinus region. The nuclear spectra, extracted from \xmm\ pn data within 11\arcsec\ of the center, are remarkably constant over the 12 years that separate the \xmm\ exposures and is consistent with no evolution of the flux or spectrum. The other two brightest point sources show strong evolution: the XRB CGX1 drops its average flux by a factor of 2.3 and the SN CGX2 evolving to lower temperatures. Observations made with resolutions worse then $\sim 5\arcmin$, will also be contaminated by the bright  ULX5 up to energies over 30~keV \citep{walton13}. Such variable sources must be taken into account to correctly interpret the spectra of different epochs.   

\section*{Acknowledgments}
We thank the anonymous referee for a thorough review and many useful suggestions that improved this paper. This work was supported under NASA Contract No. NNG08FD60C, and made use of data from the {\it NuSTAR} mission, a project led by the California Institute of Technology, managed by the Jet Propulsion Laboratory, and funded by the National Aeronautics and Space Administration. We thank the {\it NuSTAR} Operations, Software and Calibration teams for support with the execution and analysis of these observations.  This research has made use of the {\it NuSTAR} Data Analysis Software (NuSTARDAS) jointly developed by the ASI Science Data Center (ASDC, Italy) and the California Institute of Technology (USA).  We acknowledge financial support from Basal-CATA PFB-06/2007 (FEB), CONICYT-Chile FONDECYT 1140304 (PA), 1141218 (FEB), 1120061 (ET) and Anillo ACT1101 (PA, FEB, ET). AC, AM and GM acknowledge the ASI-INAF grant I/037/12/0. WNB and BL acknowledge support from  Caltech NuSTAR subcontract 44A-1092750 and NASA ADP grant NNX10AC99G. MK gratefully acknowledges support from Swiss National Science Foundation Grant PP00P2\_138979/1. PG thanks STFC for support (grant reference ST/J003697/1).

\label{lastpage}

\end{document}